\documentclass[aps,prd,twocolumn,superscriptaddress,preprintnumbers,nofootinbib]{revtex4-2}
\usepackage{amsmath}
\usepackage{amssymb}
\usepackage{natbib}
\usepackage{graphicx}
\usepackage{epsf}
\usepackage{subfigure}
\usepackage{colortbl}
\usepackage[dvipsnames, table]{xcolor}
\usepackage{threeparttable}
\usepackage{dcolumn}
\usepackage{textcomp}
\usepackage{comment}
\usepackage{epsfig}
\usepackage{xspace}
\usepackage{multirow}
\usepackage{makecell}
\usepackage{mathtools}
\usepackage{appendix}
\usepackage{booktabs}
\usepackage[utf8]{inputenc}
\usepackage[TS1,T1]{fontenc}
\usepackage{diagbox}
\usepackage{ulem}
\usepackage[
colorlinks=True,
allcolors=blue,
linktocpage=true,
]{hyperref}
\usepackage[capitalise]{cleveref}
\usepackage{latexsym}
\usepackage{array}
\usepackage{ragged2e}
\usepackage{orcidlink}
\usepackage[modulo]{lineno}
\usepackage{listings}
\usepackage{float}
\newcolumntype{C}[1]{>{\centering\let\newline\\\arraybackslash\hspace{0pt}}m{#1}}

\date{\today}

\DeclareGraphicsExtensions{.jpg,.pdf,.png,.eps,.ps}

\newcommand*{\pp}{\ensuremath{\phi\phi}}
\newcommand*{\PP}{\pp}

\newcommand*{\lcdm}{$\Lambda$CDM}
\newcommand*{\planck}{\textit{Planck}}

\newcommand{\ombh}{\ensuremath{\Omega_{\rm b} h^2}}
\newcommand{\omch}{\ensuremath{\Omega_{\rm c} h^2}}
\newcommand{\omegam}{\ensuremath{\Omega_{\rm m}}}
\newcommand{\As}{\ensuremath{A_{\mathrm{s}}}}
\newcommand{\logA}{\ensuremath{\log(10^{10}\,\As{})}}
\newcommand{\Hubble}{\ensuremath{H_0}}
\newcommand{\ns}{\ensuremath{n_\mathrm{s}}}

\newcommand*{\curv}{\ensuremath{\Omega_{\mathrm{K}}}}
\newcommand*{\kmsmpc}{\ensuremath{\mathrm{km\,s^{-1}\,Mpc^{-1}}}}
\newcommand*{\mpc}{\ensuremath{\mathrm{Mpc}}}

\newcommand{\RNum}[1]{\uppercase\expandafter{\romannumeral #1\relax}}

\newcommand{\healpix}{\texttt{HEALPix}}
\newcommand*{\polspice}{\texttt{Polspice}}
\newcommand*{\agora}{\textsc{Agora}}
\def\camb{\texttt{CAMB}\xspace}
\def\class{\texttt{CLASS}\xspace}
\def\cosmopower{\texttt{CosmoPower}\xspace}
\def\cobaya{\texttt{Cobaya}\xspace}
\newcommand{\candl}{\texttt{candl}}

\newcommand*{\ghz}{\text{GHz}}

\newcommand*{\sqdeg}{\ensuremath{{\rm deg}^2}}

\defcitealias{dutcher21}{D21}
\defcitealias{riess20}{R20}

\definecolor{amber}{rgb}{1.0, 0.49, 0.0}

\newcommand{\skipt}[1]{}
\definecolor{dblue}{RGB}{26, 158, 145}

\newcommand{\LCDM}{\lcdm}
\newcommand{\seight}{\ensuremath{\sigma_{\mathrm{8}}}} 
\newcommand{\rdrag}{\ensuremath{r_\mathrm{drag}}}

\newcommand{\Ejoint}{$\text{Ext}10\text{k}_\mathrm{joint}$}
\newcommand{\Eseparate}{$\text{Ext}10\text{k}_\mathrm{sep}$}
\newcommand{\EseparateF}{$\text{Ext}10\text{k}_\mathrm{sep}^\mathrm{FD}$}

\def\uk{\ensuremath{\mu \mathrm{K}}}

\def\ukarcmin{\uk{\rm -arcmin}}
\def\tens#1{\ensuremath{\mathsf{#1}}}
\def\arcmin{{\rm arcmin}}
\newcommand{\ext}{\text{Ext}-10\text{k}}

\newcommand*{\ellmin}{\ensuremath{\ell_{\rm min}}}
\newcommand*{\ellmax}{\ensuremath{\ell_{\rm max}}}
\newcommand*{\fsky}{\ensuremath{f_{\rm sky}}}

\begin{document}

\title{Towards constraining cosmological parameters with SPT-3G observations of 25\% of the sky}

\affiliation{Sorbonne Universit\'e, CNRS, UMR 7095, Institut d'Astrophysique de Paris, 98 bis bd Arago, 75014 Paris, France}
\affiliation{Department of Physics, University of Chicago, 5640 South Ellis Avenue, Chicago, IL, 60637, USA}
\affiliation{Kavli Institute for Cosmological Physics, University of Chicago, 5640 South Ellis Avenue, Chicago, IL, 60637, USA}
\affiliation{High-Energy Physics Division, Argonne National Laboratory, 9700 South Cass Avenue, Lemont, IL, 60439, USA}
\affiliation{Department of Physics \& Astronomy, University of California, One Shields Avenue, Davis, CA 95616, USA}
\affiliation{Fermi National Accelerator Laboratory, MS209, P.O. Box 500, Batavia, IL, 60510, USA}
\affiliation{Department of Astronomy and Astrophysics, University of Chicago, 5640 South Ellis Avenue, Chicago, IL, 60637, USA}
\affiliation{School of Physics, University of Melbourne, Parkville, VIC 3010, Australia}
\affiliation{Department of Physics and Astronomy, University of New Mexico, Albuquerque, NM, 87131, USA}
\affiliation{Kavli Institute for Particle Astrophysics and Cosmology, Stanford University, 452 Lomita Mall, Stanford, CA, 94305, USA}
\affiliation{Department of Physics, Stanford University, 382 Via Pueblo Mall, Stanford, CA, 94305, USA}
\affiliation{SLAC National Accelerator Laboratory, 2575 Sand Hill Road, Menlo Park, CA, 94025, USA}
\affiliation{University Observatory, Faculty of Physics, Ludwig-Maximilians-Universit{\"a}t, Scheinerstr.~1, 81679 Munich, Germany}
\affiliation{Enrico Fermi Institute, University of Chicago, 5640 South Ellis Avenue, Chicago, IL, 60637, USA}
\affiliation{Universit\'e de Gen\`eve, D\'epartement de Physique Th\'eorique, 24 Quai Ansermet, CH-1211 Gen\`eve 4, Switzerland}
\affiliation{Department of Physics \& Astronomy, University of Sussex, Brighton BN1 9QH, UK}
\affiliation{National Taiwan University, No. 1, Sec. 4, Roosevelt Road, Taipei 106319, Taiwan}
\affiliation{Department of Physics, University of California, Berkeley, CA, 94720, USA}
\affiliation{Universit\'e Paris-Saclay, Universit\'e Paris Cit\'e, CEA, CNRS, AIM, 91191, Gif-sur-Yvette, France}
\affiliation{Department of Astronomy, University of Illinois Urbana-Champaign, 1002 West Green Street, Urbana, IL, 61801, USA}
\affiliation{High Energy Accelerator Research Organization (KEK), Tsukuba, Ibaraki 305-0801, Japan}
\affiliation{Department of Physics and McGill Space Institute, McGill University, 3600 Rue University, Montreal, Quebec H3A 2T8, Canada}
\affiliation{Canadian Institute for Advanced Research, CIFAR Program in Gravity and the Extreme Universe, Toronto, ON, M5G 1Z8, Canada}
\affiliation{Joseph Henry Laboratories of Physics, Jadwin Hall, Princeton University, Princeton, NJ 08544, USA}
\affiliation{Department of Astrophysical and Planetary Sciences, University of Colorado, Boulder, CO, 80309, USA}
\affiliation{Department of Astronomy, University of Science and Technology of China, Hefei 230026, China}
\affiliation{School of Astronomy and Space Science, University of Science and Technology of China, Hefei 230026}
\affiliation{Department of Physics, University of Illinois Urbana-Champaign, 1110 West Green Street, Urbana, IL, 61801, USA}
\affiliation{Department of Physics and Astronomy, University of California, Los Angeles, CA, 90095, USA}
\affiliation{Department of Physics and Astronomy, Michigan State University, East Lansing, MI 48824, USA}
\affiliation{California Institute of Technology, 1200 East California Boulevard., Pasadena, CA, 91125, USA}
\affiliation{Department of Physics and Astronomy, Northwestern University, 633 Clark St, Evanston, IL, 60208, USA}
\affiliation{CASA, Department of Astrophysical and Planetary Sciences, University of Colorado, Boulder, CO, 80309, USA }
\affiliation{Department of Physics, University of Colorado, Boulder, CO, 80309, USA}
\affiliation{Department of Astronomy, Cornell University, Ithaca, NY 14853, USA}
\affiliation{Department of Physics, Case Western Reserve University, Cleveland, OH, 44106, USA}
\affiliation{Department of Physics \& Astronomy, Box 41051, Texas Tech University, Lubbock TX 79409-1051, USA}
\affiliation{Center for AstroPhysical Surveys, National Center for Supercomputing Applications, Urbana, IL, 61801, USA}
\affiliation{Dunlap Institute for Astronomy \& Astrophysics, University of Toronto, 50 St. George Street, Toronto, ON, M5S 3H4, Canada}
\affiliation{David A. Dunlap Department of Astronomy \& Astrophysics, University of Toronto, 50 St. George Street, Toronto, ON, M5S 3H4, Canada}
\affiliation{NSF-Simons AI Institute for the Sky (SkAI), 172 East Chestnut Street, Chicago, IL 60611, USA}
\affiliation{Center for Astrophysics \textbar{} Harvard \& Smithsonian, 60 Garden Street, Cambridge, MA, 02138, USA}
\author{A.~Vitrier\,\orcidlink{0009-0009-3168-092X}}
\affiliation{Sorbonne Universit\'e, CNRS, UMR 7095, Institut d'Astrophysique de Paris, 98 bis bd Arago, 75014 Paris, France}
\author{K.~Fichman}
\affiliation{Department of Physics, University of Chicago, 5640 South Ellis Avenue, Chicago, IL, 60637, USA}
\affiliation{Kavli Institute for Cosmological Physics, University of Chicago, 5640 South Ellis Avenue, Chicago, IL, 60637, USA}
\affiliation{High-Energy Physics Division, Argonne National Laboratory, 9700 South Cass Avenue, Lemont, IL, 60439, USA}
\author{L.~Balkenhol\,\orcidlink{0000-0001-6899-1873}}
\affiliation{Sorbonne Universit\'e, CNRS, UMR 7095, Institut d'Astrophysique de Paris, 98 bis bd Arago, 75014 Paris, France}
\author{E.~Camphuis\,\orcidlink{0000-0003-3483-8461}}
\affiliation{Sorbonne Universit\'e, CNRS, UMR 7095, Institut d'Astrophysique de Paris, 98 bis bd Arago, 75014 Paris, France}
\author{F.~Guidi\,\orcidlink{0000-0001-7593-3962}}
\affiliation{Department of Physics \& Astronomy, University of California, One Shields Avenue, Davis, CA 95616, USA}
\affiliation{Sorbonne Universit\'e, CNRS, UMR 7095, Institut d'Astrophysique de Paris, 98 bis bd Arago, 75014 Paris, France}
\author{A.~R.~Khalife\,\orcidlink{0000-0002-8388-4950}}
\affiliation{Sorbonne Universit\'e, CNRS, UMR 7095, Institut d'Astrophysique de Paris, 98 bis bd Arago, 75014 Paris, France}
\author{A.~J.~Anderson\,\orcidlink{0000-0002-4435-4623}}
\affiliation{Fermi National Accelerator Laboratory, MS209, P.O. Box 500, Batavia, IL, 60510, USA}
\affiliation{Kavli Institute for Cosmological Physics, University of Chicago, 5640 South Ellis Avenue, Chicago, IL, 60637, USA}
\affiliation{Department of Astronomy and Astrophysics, University of Chicago, 5640 South Ellis Avenue, Chicago, IL, 60637, USA}
\author{B.~Ansarinejad}
\affiliation{School of Physics, University of Melbourne, Parkville, VIC 3010, Australia}
\author{M.~Archipley\,\orcidlink{0000-0002-0517-9842}}
\affiliation{Department of Astronomy and Astrophysics, University of Chicago, 5640 South Ellis Avenue, Chicago, IL, 60637, USA}
\affiliation{Kavli Institute for Cosmological Physics, University of Chicago, 5640 South Ellis Avenue, Chicago, IL, 60637, USA}
\author{D.~R.~Barron\,\orcidlink{0000-0002-1623-5651}}
\affiliation{Department of Physics and Astronomy, University of New Mexico, Albuquerque, NM, 87131, USA}
% ...
\collaboration{SPT-3G Collaboration}
\noaffiliation  
\begin{abstract}
The South Pole Telescope (SPT), using its third-generation camera, SPT-3G, is conducting observations of the cosmic microwave background (CMB) in temperature and polarization across approximately $10\,000\,\sqdeg$ of the sky at 95, 150, and 220 GHz. This comprehensive dataset should yield stringent constraints on cosmological parameters.
In this work, we explore its potential to address the Hubble tension by forecasting constraints from temperature, polarization, and CMB lensing on early dark energy (EDE) and the variation in electron mass in spatially flat and curved universes.
For this purpose, we investigate first whether analyzing the distinct SPT-3G observation fields independently, as opposed to as a single, unified region, results in a loss of information relevant to cosmological parameter estimation. We develop a realistic temperature and polarization likelihood pipeline capable of analyzing these fields in these two ways, and subsequently forecast constraints on cosmological parameters. Our findings indicate that any loss of constraining power from analyzing the fields separately is primarily concentrated at low multipoles ($\ell < 50$) 
and the overall impact on the relative uncertainty on standard $\Lambda$ cold dark matter parameters is minimal ($<3\%$). 
Our forecasts suggest that SPT-3G data should improve by more than a factor of 90 and 190 the figure of merit of the EDE and the varying electron mass models, respectively, when combined with \planck\ data. The likelihood pipeline developed and used in this work is made publicly available online.

\end{abstract}


\maketitle
\tableofcontents
\section{Introduction}

Over the last 30 years, observations of the cosmic microwave background (CMB) have steadily driven advances in our understanding of the Universe. The statistical properties of the CMB, as revealed by the European Space Agency's \planck\ satellite and the latest measurements from the South Pole Telescope (SPT) \cite{Ge25,Camphuis25} and the Atacama Cosmology Telescope \cite{louis25}, are all consistent with the standard $\Lambda$ cold dark matter (\LCDM) cosmological model, and constrain most of its parameters at better than the percent level \citep{planck18-6,louis25,Camphuis25}. Despite these successes, old and new questions remain unresolved. These include the nature of dark matter and dark energy, as well as the origin of primordial perturbations. More recently, the precision of the current data is leading to tensions in some parameter values. The underlying causes of the Hubble tension \citep{breuval_small_2024} and the emerging tension between baryon acoustic oscillation (BAO) and CMB data \citep{desi25, Ferreira25} are unclear regarding the extent to which they may be due to new physics or unknown systematic errors.

Ongoing and planned CMB experiments aim to address these questions by unveiling new information in the CMB temperature anisotropies at small angular scales and in the polarization of the CMB at all angular scales. These include the Simons Observatory \cite{SimonsObs25} and the ongoing set of surveys with the third-generation camera on the SPT, SPT-3G \citep{benson14,sobrin22,prabhu24}. SPT-3G deep observations will also enable delensing of degree-scale B-mode polarization maps from the BICEP/Keck family of telescopes, which will sharpen the search for primordial gravitational waves from inflation \citep{bicepkeck2022}.

In total, SPT-3G has surveyed 25\% of the sky in frequency bands centered at 95, 150, and $220\,\ghz$, with coadded sensitivity between three and 17 times higher than \planck\footnote{The \planck\ 2018 data have temperature white noise levels of 77, 33, and $47\,\ukarcmin$ at 100, 143, and $217\,\ghz$, corresponding to a coadded sensitivity of $27.6\,\ukarcmin$.} and with five times better resolution. The SPT-3G observations have been primarily divided into three surveys: the \textit{Main} field, observed for five Austral winter seasons, the \textit{Summer} fields, observed for four Austral summer seasons, and the \textit{Wide} fields, observed for one entire year.

We define the combined SPT-3G surveys, covering a total of $10\,000\,\sqdeg$, as \ext.  The \ext\ SPT-3G survey is anticipated to yield stringent constraints on cosmological parameters. Notably, Ref.~\cite{prabhu24} forecasted that \ext\ observations could improve constraints on certain cosmological parameters within both \LCDM\ and extended models up to a factor of 2 compared to \planck\ results and up to a factor of 3 when combined with \planck\ data. These forecasts, however, assumed the \textit{Main}, \textit{Summer}, and \textit{Wide} fields to be five aggregated sky patches, one for the \textit{Main} field, three for the \textit{Summer} fields, and one for the \textit{Wide} fields. However, in reality the \textit{Wide} fields consist of nine distinct regions. Thus, in total, the \ext\ survey is composed of 13 individual regions which exhibit unique observational characteristics, including variations in observation elevation, atmospheric contamination, Galactic foreground contamination, and noise properties. To accurately account for these specificities, the most straightforward approach is to treat all fields independently. 

In this paper, we investigate whether this independent analysis approach allows us to fully capture the cosmological information encoded in the observed sky.
Specifically, we address the question of whether analyzing the fields separately diminishes their constraining power by neglecting the correlations between pixels across different fields, which also impacts access to larger angular scale modes. To answer this question, we compare the constraints on \LCDM\ cosmological parameters obtained from two distinct analysis approaches: one treating the entire survey as a single, contiguous field, and another treating it as 13 separate fields. To isolate the impact of the neglected correlations, we perform this comparison assuming identical noise levels across all fields in both analysis scenarios, specifically adopting the noise properties characteristic of the \textit{Wide} fields.


We perform this investigation for the CMB power spectra in temperature (TT), polarization (EE), and the cross-correlation between the two (TE).
We find that the impact on \LCDM\ cosmological parameter constraints is less than $3\%$. We thus adopt the separate-field approach to build mock likelihoods for the full-depth \textit{Main}, \textit{Summer}, and \textit{Wide} fields using their own proper noise, and forecast cosmological parameters. We confirm the findings of \cite{prabhu24} in \LCDM\ and extend the forecasting work on models proposed to solve the Hubble tension, as in \cite{khalife24}. For this part, we also develop a lensing reconstruction mock likelihood. 

The paper is organized as follows: in Sec.~\ref{sec:SPT} we describe the \ext\ survey. Section~\ref{sec:likelihood} then details how we build the likelihoods and Fisher matrices to forecast constraints. We present our results in Sec.~\ref{sec:results} and we summarize our conclusions in Sec.~\ref{sec:conclusions}.

\section{SPT-3G and the \ext\ survey footprint}
\label{sec:SPT}

The SPT is a 10-meter diameter telescope \cite{carlstrom11} located at the Amundsen-Scott South Pole station.  The telescope has arcminute angular resolution and is  currently equipped with its third-generation survey camera, SPT-3G. The SPT-3G focal plane consists of approximately 16\,000 polarization-sensitive detectors that observe in three frequency bands centered at 95, 150, and 220 GHz \cite{benson14, sobrin22}.

\begin{figure*}
\includegraphics[width=1\textwidth]{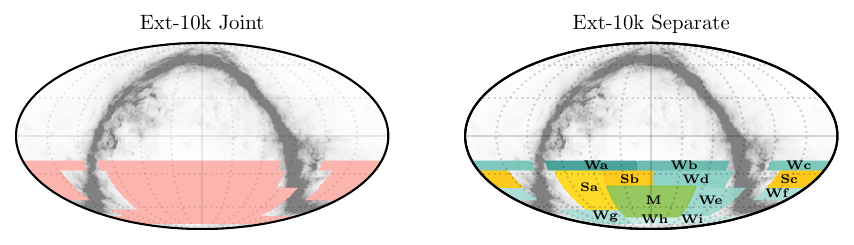}
\caption{Observed sky area with joint (left) and separated (right) masks used to perform either a global or a separate analysis of the \ext\ survey, which spans declinations from $-20^\circ$ to $-80^\circ$. The \textit{Main} field (in green) is denoted by M, the three \textit{Summer} fields (in yellow) are Sa, Sb, and Sc, and the nine \textit{Wide} fields (in blue) are named from Wa to Wi. The masks are shown in equatorial coordinates and are superposed on the \texttt{Commander} dust map produced in the component separation of the \planck\ analysis \citep{planck18-4}. The dashed lines are $30^\circ$ intervals between meridians and between parallels. The vertical and horizontal solid lines correspond to $\text{RA}=0$ and $\text{dec}=0$, respectively.}
\label{fig_2ways}
\end{figure*}

\begin{table*}[]
    \centering
    \begin{tabular}{l|>{\Centering}p{0.8cm}|>{\Centering}p{0.8cm} >{\Centering}p{0.8cm} >{\Centering}p{0.8cm}|>{\Centering}p{0.8cm} >{\Centering}p{0.8cm} >{\Centering}p{0.8cm} >{\Centering}p{0.8cm} >{\Centering}p{0.8cm} >{\Centering}p{0.8cm} >{\Centering}p{0.8cm} >{\Centering}p{0.8cm} >{\Centering}p{0.8cm}}
    \hline
    \hline
    \multirow{2}{*}{} & \multirow{2}{*}{\textit{Main}} & \multicolumn{3}{c|}{\textit{Summer}} & \multicolumn{9}{ c}{\textit{Wide}} \\
    & & a & b & c & a & b & c & d & e & f & g & h & i \\
    \hline
    \hline
    \fsky[\%]\ & 4.04 & 2.92 & 1.31 & 1.94 & 1.28 & 1.28 & 1.67 & 2.04 & 2.03 & 1.22 & 1.12 & 0.65 & 0.65  \\
    \hline
    Coadded sensitivity [$\ukarcmin$]& 1.9 & 6.2 & 6.8 & 6.6 & \multicolumn{9}{ c}{8.8}  \\
    \hline
    \end{tabular}
    \caption{Fraction of the total sky, \fsky, and coadded sensitivities of the 13 fields of the \ext\ survey. The \fsky\ values are based on the masks used in the \Eseparate\ analysis. The \textit{Main} field is the biggest field and covers 4\% of the total sky, while the \textit{Wide} h and i fields each cover only 0.65\%. The \textit{Summer} fields and the \textit{Wide} fields cover 6\% and 12\% of the sky, respectively, after apodizing each field individually.}
    \label{tab_fsky}
\end{table*}

SPT-3G has been used to survey approximately $10\,000\,\sqdeg$ of the Southern sky, providing a powerful dataset for cosmology. The survey is partitioned into different observational fields, each optimized to fulfill different scientific objectives and operational constraints. The fields are defined as follows:

\textbf{\textit{Main}}: During the Austral winter, spanning roughly eight months annually, SPT-3G has primarily been used to observe a $1500\,\sqdeg$ \textit{Main} field.  This field has the lowest Galactic foreground contamination. From data obtained between 2019 and 2023, the field has been measured to a coadded sensitivity of approximately $1.9\,\ukarcmin$, the deepest of any of the SPT-3G fields. The \textit{Main} field will be observed for at least two more winter seasons, further improving the coadded sensitivity to $<1.6\,\ukarcmin$. The deep integration within this region enables a precise measurement of polarization E-modes. This measurement and the reconstructed lensing potential will be used to delens the data of the BICEP/Keck experiment \cite{bicepkeck2022} and to set tight constraints on the tensor-to-scalar ratio \cite{bicepkeck2022,Zebrowski25}.

\textbf{\textit{Summer}}: During the Austral summer, starting in early December through the middle of March, the Sun contaminates the \textit{Main} field. During this period between 2019 and 2023, SPT-3G 
observed three distinct \textit{Summer} fields totaling approximately $2600\,\sqdeg$ to a coadded sensitivity of approximately $6\,\ukarcmin$. Although less deep than the \textit{Main} field, the expanded sky coverage reduces sample variance in CMB power spectra measurements on large angular scales, thereby 
improving overall cosmological parameter constraints.

\textbf{\textit{Wide}}: Between 2023 and 2024, SPT-3G observed nine fields covering an additional $6000\,\sqdeg$ in what we call the \textit{Wide} survey. This survey targets all sky regions accessible from the South Pole within instrumentally feasible elevations, ranging from approximately $20^\circ$ to $80^\circ$, while avoiding the Galactic plane.\footnote{\textit{Wide} e partially scans through the Galactic plane for simplicity of definition of survey boundaries.}
The \textit{Wide} fields are the shallowest of the 
three surveys, measured to a coadded sensitivity of approximately $9\,\ukarcmin$. However, the additional area helps to further reduce sample variance of the combined SPT-3G \ext\ survey measurements. 
The nine \textit{Wide} fields were defined and observed to minimize contamination from the Moon, Sun, and atmosphere, while still maintaining a high observing efficiency. 
At low elevations, the Moon passes through some observation fields. 
Field observations were scheduled to avoid Moon contamination over the lunar cycle. Similarly, during the Austral summer when the Sun is above the horizon, only fields through which the Sun does not pass are observed. To minimize atmospheric contamination, low-elevation fields were observed during the Austral winter, while high-elevation fields were observed during the Austral summer.

The footprints of the \textit{Main}, \textit{Summer}, and \textit{Wide} fields are shown in Fig.~\ref{fig_2ways}. 
The masks used in this work are produced in \healpix\footnote{\url{https://healpix.sourceforge.io}} \citep{gorski05} 
at $N_{\text{side}}=2048$.\footnote{See Sec.~4 of \citep{gorski05} for the definition of $N_{\text{side}}$.} The binary masks are then apodized to reduce ringing in harmonic space for power spectrum estimation and to reduce correlations between modes. We apodize with the equivalent of a Gaussian taper of width $\sigma=34\,\arcmin$. The apodization algorithm is the same as the one described in Appendix~A of \cite{planck15-11}. Table~\ref{tab_fsky} summarizes the \fsky\ values for the 13 different fields of the survey.

While the \textit{Wide} field observations are designed to avoid the Galactic plane, some highly emissive Galactic regions are still included in the survey and will be masked in the real analysis. In this work, we do not account for the masking of these contaminated regions but we verify that applying the \texttt{GAL080} mask produced by \planck\ collaboration \citep{planck15-13, planck15-11}, which removes the 20\% of the full sky most contaminated by Galactic dust emission, only changes the \textit{Wide} field coverage from 14.4\% to 13.2\% of the total sky.

In the following sections, we will consider two ways of analyzing the \ext\ survey. In the first method, which we call \Ejoint, we consider the 13 fields as a single patch surrounded by a common apodized border and covering 23\% of the sky. In the second one, which we call \Eseparate, we consider the 13 separate fields of the \ext\ survey, each having apodized borders, and covering 22\% of the sky. The small difference between sky fractions is due to the loss of sky area due to the apodization of the 13 separate fields in the \Eseparate\ case compared to the single field in the \Ejoint\ case. Figure~\ref{fig_2ways} shows the two cases.

\section{Methodology}
\label{sec:likelihood}
\subsection{Likelihood}
The realistic likelihoods we build to forecast the constraints on cosmological parameters in different settings are implemented in the \texttt{JAX}-friendly \citep{jax2018github} fully differentiable CMB power spectra likelihood framework \candl\footnote{\url{https://github.com/Lbalkenhol/candl}} \cite{balkenhol24}, which allows us to automatically estimate Fisher matrices when coupled to the differentiable emulator \cosmopower\ \citep{spuriomancini22,piras23}. We build the likelihood of the mock data given a model following the SPT-3G 2018 \cite{balkenhol23} likelihood, as implemented in \candl. Our analysis being at high $\ell$ with a sufficient number of independent modes, we assume the power spectrum amplitudes are Gaussian distributed. The likelihood thus takes the form
\begin{equation}
-\ln{\cal L}(\vec{\hat D }| \vec{D}(\vec{\theta})) = \frac{1}{2} \left[\vec{\hat D} - \vec{D}(\vec{\theta})\right]^{\tens{T}} \Sigma^{-1}
 \left[\vec{\hat D} - \vec{D}(\vec{\theta})\right] + {\rm const},
 \label{eq:basic-likelihood}
 \end{equation}
where $\vec{\hat D}$ is the data vector, $\vec{D}(\vec{\theta})$ is the model with parameters $\vec{\theta}$, and $\Sigma$ is the covariance matrix.
 
\subsection{Forecasts} 
We derive constraints on parameters either using a Fisher matrix formalism for the \ext\ TT/TE/EE analysis strategy validation, or a full Markov chain Monte Carlo (MCMC) for the \ext\ \LCDM\ and extended models' forecasts. The use of MCMC is necessary to explore highly non-Gaussian posterior distributions, such as those of the extended models (see Fig.~\ref{fig_VarMeOmk} as an example). For the calculation of the theory CMB power spectra, we either use the Boltzmann solvers \class\ \citep{lesgourgues11} or \camb\ \citep{lewis11b} or the emulator \cosmopower. For our forecasts, we adopt the same nuisance parameters as assumed in \cite{balkenhol23} (see Table~VIII from that work). Finally, we use a Gaussian prior on $\tau$ with 0.054 as the central value and 0.0074 as the standard deviation $\mathcal{N}\sim(0.054,0.0074)$, based on the \planck\ 2018 results \cite{planck18-6}. 

The Fisher matrix is calculated from a second-order Taylor expansion of the likelihood around the best-fit parameters and allows us to forecast cosmological parameter constraints without any data. It depends on the power spectrum covariance matrix and on the theoretical power spectra as follows: 
\begin{equation}
    \mathcal{F}_{\alpha \beta}=\sum_{b_1,b_2}\sum_{XY,WZ}\frac{\partial D_{b_1}^{XY}}{\partial \theta^{\alpha}}\left[\text{Cov}(D_{b_1}^{XY},D_{b_2}^{WZ})\right]^{-1}\frac{\partial D_{b_2}^{WZ}}{\partial \theta^{\beta}},
\end{equation}
where $\{XY,WZ\} \in \{\text{TT},\text{TE},\text{EE}\}$, $\theta^{\alpha}$ refers to the cosmological and nuisance parameters, and the indices $b_1$ and $b_2$ refer to the binned power spectrum estimates. 
Thanks to our differentiable pipeline built on \cosmopower-\texttt{JAX} \cite{spuriomancini22,piras23} and \candl, we can get accurate derivatives quickly using \texttt{JAX}. This allows us to calculate Fisher matrices avoiding numerical difficulties associated with the finite difference method.


\subsection{Covariance matrix}
\label{subsection_covariance}
We calculate a realistic TT/TE/EE covariance matrix including several real data effects, as in \cite{Camphuis25}.
We use an analytical calculation using the code developed in \cite{Camphuis22} and using a narrow kernel 
approximation (NKA) \cite{Efstathiou04,GarciaGarcia19} to speed up the computations. The computation correctly accounts for the mode coupling 
introduced by the apodized footprint masks. In particular, it includes the effect of debiasing the pseudo-$C_{\ell}$s from the effect of the mask using \polspice\ \citep{hivon02,chon04}, as described in Sec.~5.4 of \cite{Camphuis22}. In this work, we do not include the effect of masking point sources in the covariance, under the assumption that the \ext\ analysis will inpaint over bright sources, as was done in \cite{Camphuis25}. For the fiducial cosmological model, the theory power spectrum is calculated from the \planck\ 2018 best fit.\footnote{\texttt{base\_plikHM\_TTTEEE\_lowl\_lowE\_lensing}} The foreground levels at each frequency are 
based on the \agora\ simulations \cite{omori22}, which include different extragalactic components such as the thermal Sunyaev–Zel’dovich (tSZ)
and kinematic Sunyaev–Zel’dovich (kSZ) effects, cosmic infrared background (CIB), and radio sources. Those simulations do not include Galactic 
dust. 

\begin{figure}
\includegraphics[width=0.5\textwidth]{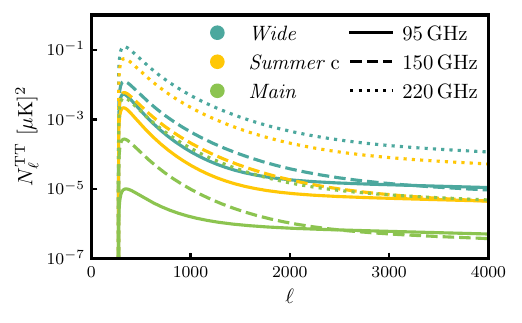}
\caption{Expected temperature noise power spectra for the \textit{Wide}, \textit{Summer} c, and \textit{Main} fields for the three autofrequencies $95\,\ghz\times95\,\ghz$, $150\,\ghz\times150\,\ghz$, and $220\,\ghz\times220\,\ghz$. They include the transfer function shown in Fig.~\ref{fig_tf} and the pixel window function. The visible suppression at low multipoles corresponds to a suppression seen both in signal and noise due to the TOD filtering.}
\label{fig_noise_curves}
\end{figure}

The noise power spectra used to build the covariance matrix are the same as the ones used and described in 
\cite{prabhu24} for the full SPT-3G survey observations taken between 2019 and 2024. Namely, the \textit{Main} field 
noise curves include the five winter seasons from 2019 to 2023, the \textit{Summer} field noise curves include four 
Austral summer seasons between 2019 and 2023, and the \textit{Wide} field noise curves correspond to one year of observation between 2023 and 2024. These noise spectra are modeled according to Eq.~(1) of \cite{prabhu24} and include instrumental and atmospheric noise. The \textit{Main} and \textit{Summer} fields' noise spectra are obtained by fitting real noise data to the model. The \textit{Wide} field noise curves have the same parametrization as the \textit{Summer} noise curves, since real noise power spectra for the \textit{Wide} fields did not exist when this work started. First preliminary measurements of the distinct \textit{Wide} fields' temperature noise spectra derived later show good agreement with the modeled noise power spectra. On top of these noise spectra we add a transfer function and a pixel window function which mimic the effect on the power spectrum of the time-ordered data (TOD) filtering and the integration of signal (or noise) over the pixel area, respectively. The TOD filtering contributes to suppressing both the signal and noise at low $\ell$. We show in Fig.~\ref{fig_noise_curves} the expected temperature noise power spectra of the \textit{Wide} fields, the \textit{Main} field, and the \textit{Summer} c field (see 
\cite{prabhu24} for details). Since \textit{Summer} a and \textit{Summer} b have comparable white noise levels as \textit{Summer} c, we only show \textit{Summer} c noise curves for comparison purposes with the \textit{Wide} and \textit{Main} fields. 

We account for the impact of deconvolving the beams, the pixel window function, and the transfer function (see Fig.~\ref{fig_tf}), the computation of which is detailed in Appendix~\ref{sec:TF}. In particular, we account for the anisotropic nature of our filtering along the scan direction by rescaling the diagonal of the covariance 
matrix, as also detailed in Appendix~\ref{sec:TF}. The diagonal rescaling estimation from simulations is detailed in Sec.~IV~F of 
\cite{Camphuis25} and in Hivon \textit{et al.}, in prep. This rescaling is denoted $H(\ell)$ and is set to be equal to the transfer function based on \cite{Camphuis25} and Hivon \textit{et al.}, in prep, where this approximation is shown to be accurate 
at the 10\% level for the \textit{Main} and \textit{Summer} fields.  
\begin{figure}
\includegraphics[width=0.5\textwidth]{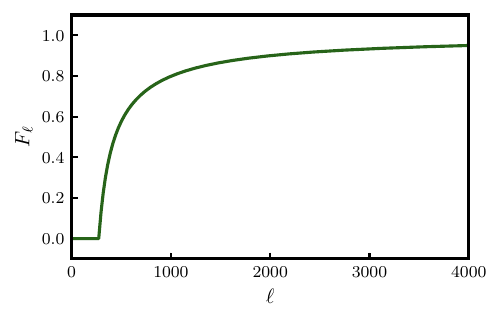}
\caption{Transfer function used in our TT/TE/EE analysis. This function describes the effect of the high-pass and low-pass filtering of the data on the power spectrum and the covariance matrix.}
\label{fig_tf}
\end{figure}

\subsection{Model}
\label{subsection_model}
Our model vector $\vec{D}(\vec{\theta})$ is described by a set of cosmological and nuisance (foreground) parameters. The six \LCDM\ cosmological parameters are the current expansion rate of the Universe \Hubble, the baryon density \ombh, the cold dark matter density \omch, the optical depth to reionization $\tau$, and the amplitude \As\ and spectral index \ns\ of the primordial scalar power spectrum. In addition to the cosmological parameters that affect the CMB theory spectra, we model the foreground parameters, as in \cite{balkenhol23}. 

We account for the Poisson-distributed unresolved radio galaxies and dusty star-forming galaxies through 12 amplitude parameters for the six different auto- and cross-frequency spectra in temperature $D^{\text{TT},\text{Poisson}}_{3000,\mu\times\nu}$ and in polarization $D^{\text{EE},\text{Poisson}}_{3000,\mu\times\nu}$, where $\{\mu,\nu\}\in\{95,150,220\}$. The Galactic dust is modeled in TT, TE, and EE by a modified black-body with an amplitude $A^{XY,\text{Dust}}_{80}$, a spectral index $\beta^{XY,\text{Dust}}$, and a power law index $\alpha^{XY,\text{Dust}}$, where $\{X,Y\}\in\{\text{T},\text{E}\}$. We assume the dust contamination is constant across fields. The clustering term of the CIB is also modeled by a modified black-body spectrum with an amplitude $A^{\text{CIB-cl}}_{3000}$ and a spectral index $\beta^{\text{CIB-cl}}_{3000}$. We account for the tSZ and kSZ effects through two amplitude parameters $A^{\text{tSZ}}$ and $A^{\text{kSZ}}$. The correlation between the tSZ and the CIB is modeled through the correlation parameter $\xi^{\text{tSZ}\times\text{CIB}}$. We also sample over supersample lensing through a parameter $\kappa$ in the same manner as \cite{dutcher21}. Although the prior on this parameter depends on the field size \cite{manzotti14} we decide to be conservative and use the \textit{Main} field value, knowing that the prior for the \ext\ real analysis will be tighter. We check that by fixing this parameter in our \LCDM\ \ext\ TT/TE/EE/\PP\ forecasts, the only significantly impacted parameter is \Hubble, with the error bar reduced by 9\%.

Our model does not account for instrumental systematics, such as the calibration and the beam. In the real analysis of one field, we sample over six calibration parameters for the three frequencies in temperature and polarization. In the real analysis of the \ext\ survey, we will sample over 78 calibration parameters. 

\subsection{Mock-data vector}
Our data vector is composed of binned simulated auto- and cross-frequency TT, TE, and EE power spectrum estimates, or bandpowers. The bin width is chosen to be $\Delta\ell=50$,
as in previous SPT-3G TT/TE/EE analyses \cite{balkenhol23,Camphuis25}, with $\ell_\mathrm{min} = 350$ and $\ell_\mathrm{max} = 4000,$ also similar to previous analyses. In each bin there are 18 total bandpowers, comprising all possible combinations of T and E and the three frequency bands (95, 150, and $220\,\ghz$). We treat TE and ET together for auto- and cross-frequencies.
We assume that the data power spectra are always calculated from cross-correlating maps from different observation times to avoid noise bias. We build our simulated data as a smooth fiducial spectrum with \candl. The theory CMB is calculated from the \planck\ 2018  
cosmological parameters (see Table~1 of \citep{planck18-6}) using either Boltzmann solvers such as \class\ and \camb\ or the \cosmopower\ emulator. For the latter, we use the SPT high-accuracy models\footnote{\url{https://github.com/alessiospuriomancini/cosmopower/tree/main/cosmopower/trained_models/SPT_high_accuracy}} (see \cite{balkenhol23} for more details). The foregrounds at each frequency are modeled as in \cite{balkenhol23}, where Table~VIII summarizes the nuisance parameter central values used to produce our data vector. 

We assume that in the real analysis the \polspice\ estimator is used to correct for the effect of the cut sky. Masking the sky suppresses the pseudo-$C_\ell$ power spectrum, as described in Eq.~(14) of \cite{hivon02}. The \polspice\ procedure corrects for this effect through an apodization function (see Appendix~\ref{apodizesigma_info}) that depends on a parameter named \texttt{apodizesigma}.
Because of this correction, the \polspice\ spectra are biased and linked to the unbiased power spectra via the \polspice\ kernel (see also Sec.~6 of \cite{Camphuis22}). We take this into account in our data vector. We use an \texttt{apodizesigma} value of $30^\circ$ for the 13 fields to maintain consistency across all fields (see Appendix~\ref{apodizesigma} for further details on this choice). We increase this value to $80^\circ$ for the \Ejoint\ analysis because it allows us to include larger angular scales than $30^\circ$. The multiplication of the \polspice\ kernel and the binning matrix forms the window functions that compress the $\ell$-by-$\ell$ spectrum into binned bandpowers, $D_b= W_{b\ell}C_\ell$. Hence, we obtain $D_b$ from $C_{\ell}$ as follows:
\begin{equation}
    D_b= Q_{b\ell'}K_{\ell'\ell}C_\ell,
\end{equation}
where $K$ is the \polspice\ kernel and $Q$ is the binning matrix which encapsulates the conversion from $C_\ell$ to $D_\ell$, as detailed in Eq.~(28) of \cite{Camphuis25}.

\subsection{Cases considered} 
\label{subsec_likelihood}
We build the likelihood for three different cases:
\begin{enumerate}
    \item The \Ejoint\ case, where we treat the entire survey as one contiguous field with homogeneous noise levels equivalent to those of the \textit{Wide} fields.
    \item The \Eseparate\ case, where we treat the entire survey as 13 separate fields with homogeneous noise levels equivalent to those of the \textit{Wide} fields.
    \item The \EseparateF\ case at full depth, where we treat the entire survey as 13 separate fields with the \textit{Main}, \textit{Summer}, and \textit{Wide} fields having their own distinct noise levels (see Table~1 of \cite{prabhu24} for the detailed white noise levels of the different fields).
\end{enumerate}
The comparison of the first two cases allows us to determine the impact of dividing the observed sky in 13 different regions, assuming the same noise levels in the two cases. The third case is the most realistic, with the proper noise curves for all the fields. We use this likelihood for the forecasts in Sec.~\ref{sec:results}. In the first case, only one covariance matrix is computed. In the second and third cases, we calculate 13 matrices. Note that in the three cases all the parameters which describe extragalactic foregrounds are assumed to be the same across all the fields. We also use one single set of parameters to describe the Galactic dust. 
However, in the third case we also test 13 different sets of parameters to capture the potentially different levels of contamination of each field. 

\subsection{Lensing likelihood}
\label{sec_lensing}
Forecasts are performed for TT/TE/EE, however, we also build a mock lensing likelihood to compare our results with \cite{prabhu24} and to forecast constraints on extended models in Sec.~\ref{sec:results_MCMC}. 
For this case, we build five separate likelihoods, one for the \textit{Main} field, three for each \textit{Summer} field, and one for the whole \textit{Wide} field. The CMB lensing analysis strategy will be investigated in a future work. The lensing data vector is composed of a normalized theory lensing potential spectrum 
$C^{\kappa\kappa}_{\textit{L}}=[L(L+1)]^2C_L^{\phi\phi} / 2\pi$ produced either by \class\ or \camb\ from the same cosmological parameters as for the TT/TE/EE data vector. We then bin the lensing data vector with a bin size $\Delta\textit{L}=10$. We use the fiducial lensing potential spectrum $C^{\phi\phi}_{\textit{L}}$ and the lensing reconstruction noise curves $N^{\phi\phi}_{\textit{L}}$ used in \cite{prabhu24} to produce the diagonal lensing analytic covariance matrices. The \textit{L} range is $30\leq\textit{L}\leq3500$ and we take into account the binning factor as follows:
\begin{equation}
    \Sigma^{\text{lensing}}_{\textit{L}\textit{L}}=\frac{2}{(2\textit{L}+1)f_{\text{sky}}\Delta\textit{L}}(C^{\kappa\kappa}_{\textit{L}}+N^{\kappa\kappa}_{\textit{L}})^2.
\end{equation}
We do not include foregrounds in the lensing data vector or covariance. Our window function is the binning matrix. 

\subsection{Combination with \planck}
\label{sec_planck}
In our forecasts, we also explore the combination of the \ext\ survey with \planck\ for \lcdm\ and extended models. In \cite{prabhu24}, it was shown that the full \ext\ survey is expected to have higher constraining power than \planck\ at $\ell\gtrsim1800$ for TT, $\ell\gtrsim800$ for TE, $\ell\gtrsim450$ for EE, and at all scales for $\phi\phi$. 

In this work, we replace the real \planck\ data vector in the TT/TE/EE high-$\ell$ \planck\ likelihood \texttt{plik\_rd12\_HM\_v22\_TTTEEE.clik} by a theory power spectrum calculated  with either \class\ or \camb\ from the same cosmological parameters used for the SPT-3G TT/TE/EE and lensing data vectors. On top of this theory CMB, we also add a fiducial foreground and instrumental contribution assuming the \planck\ 2018 best fit for the nuisance parameters.\footnote{\texttt{base\_plikHM\_TTTEEE\_lowl\_lowE\_lensing}} 
We perform the same $\ell$-cuts as \cite{prabhu24} in TT/TE/EE likelihoods to avoid double counting information, namely for the \ext\ likelihoods we keep $\ell>2000$ for TT, $\ell>1000$ for TE, and $\ell>750$ for EE. To apply these cuts, we use \texttt{clipy},\footnote{\url{https://github.com/benabed/clipy}} a \texttt{python} implementation of the \planck\ data with \texttt{JAX} support. We do not use the \planck\ lensing likelihood and we keep all the scales in our SPT-3G \ext\ lensing likelihoods. Moreover, we do not use \planck\ low-$\ell$ EE likelihood but set the same $\tau$ prior $\mathcal{N}\sim(0.054,0.0074)$ as in our forecasts without \planck. When exploring parameters, we assume that priors on the \planck\ nuisance parameters are the same as in the \planck\ 2018 analysis (see Table~16 of \cite{planck18-5}).

\section{Results}
\label{sec:results}
\label{results}
\subsection{Analysis strategy validation}

In this section, we compare the constraints from the \Ejoint\ and \Eseparate\ cases, using the same \textit{Wide} field noise curves for both across the whole footprint. 
We include TT, TE, and EE bandpowers between multipoles $\ellmin=350$ and $\ellmax=4000$. The \ellmin\ is chosen to cut 
out the scales suppressed by the filtering of the TOD, and it is similar to the choice adopted for the real 
data \cite{Camphuis25}. 

We first compare the bandpower covariance matrices between the \Ejoint\ and \Eseparate\ cases. The ratio of the diagonal elements of the two matrices is shown in Fig.~\ref{fig_Cl_cov_ratio_binned400}. We detail the computation of the \Eseparate\ bandpower covariance matrix in Appendix~\ref{sec:coadded_cov}. We show the ratio of the TT and EE $95\,\ghz\times95\,\ghz$ diagonals after binning the covariance in $\Delta\ell=400$ to decrease the bin-to-bin correlations so that the full constraining power is contained in the diagonal. The two cases differ by $~5\%$. We obtain the same result for all the auto- and cross-frequency TT, TE, and EE diagonals. This difference corresponds to the decrease in sky fraction due to the additional apodized borders of the 13 fields in the \Eseparate\ case (the \Ejoint\ case has $\fsky = 0.2322$, while the \Eseparate\ case has $\fsky = 0.2215$, with a relative difference of $4.8\%$). Apart from this, we do not detect a significant difference between the two cases due to the loss of cross-field modes in the \Eseparate\ case. 
\begin{figure}
\includegraphics[width=0.5\textwidth]{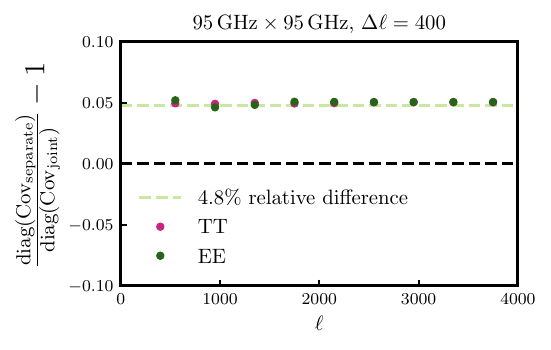}
\caption{Relative difference in the diagonal of the bandpower covariance matrix between \ext\ separate and joint analyses. We show TT and EE $95\,\ghz\times95\,\ghz$ autofrequency with a bin size $\Delta\ell=400$ for $\ell_{\text{min}}=350$. The two diagonals differ by $\sim5\%$ which corresponds to the 4.8\% difference in sky fraction between the two cases due to the apodization of the individual masks in the \Eseparate\ analysis.}
\label{fig_Cl_cov_ratio_binned400}
\end{figure}

\begin{figure*}
\includegraphics[width=1\textwidth]{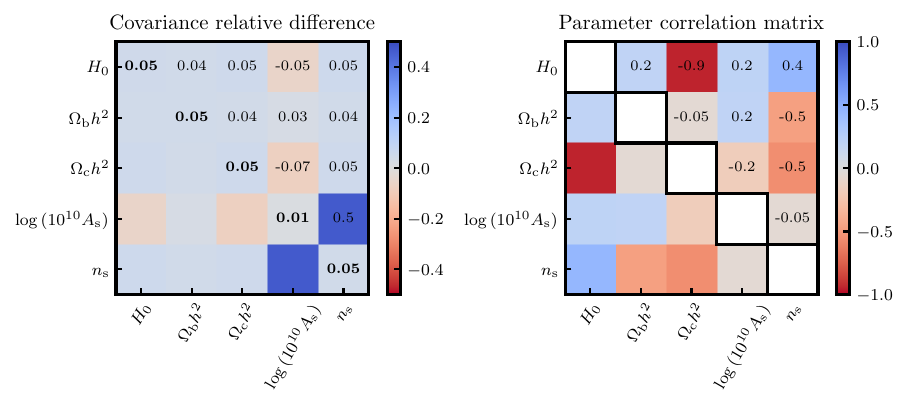}
\caption{Left: relative difference in the parameter covariance matrix between \ext\ separate and joint analyses. Analyzing the fields individually contributes to a $\sim5\%$ increase of parameter variance. We do not show $\tau$ since it is prior driven. Right: parameter correlation matrix for the \ext\ separate analysis obtained from Fisher forecasting.}
\label{fig_ratio_param_cov}
\end{figure*}

We then compare the Fisher matrices calculated in the two cases for the \LCDM\ model. This comparison is to verify how much the loss of \fsky\ affects constraints on parameters. In the \Ejoint\ analysis, we compute one Fisher matrix using the corresponding NKA covariance matrix, while in the \Eseparate\ analysis 13 Fisher matrices are obtained from the 13 NKA covariance matrices. We then sum the Fisher matrices together in order to obtain the final Fisher matrix of the separate analysis,
\begin{equation}
    \mathcal{F}^{\text{sep}}=\sum_i \mathcal{F}^i.
\end{equation}
The left diagram of Fig.~\ref{fig_ratio_param_cov} shows the relative difference between the parameter covariance matrices obtained in the \Eseparate\ and \Ejoint\ cases. One can see that the differences in the diagonal, namely in the variance of each parameter, are small, of the order of $\sim 5\%$. These differences in parameter variance translate into differences of the order of $\sim 3\%$ in the resulting error bars, which is smaller than the numerical precision in parameter estimation with an MCMC \cite{Lewis_getdist}. The largest relative difference is in the covariance between \logA\ and \ns, which is 50\%. However, the absolute correlation value in the two cases is only $-0.05$ and $-0.03$ in the \Eseparate\ and \Ejoint\ cases, respectively. An increase of 50\% of a small value has thus a negligible impact on the overall constraining power. The degeneracy between these parameters is particularly sensitive to how extended the range of well-measured multipoles is. Since in the \Ejoint\ case the power spectrum error bars at all multipoles are marginally more constraining due to the slightly larger \fsky, the degeneracy between the two parameters can more effectively be lifted compared to the \Eseparate\ case.
We also show on the right-hand side of Fig.~\ref{fig_ratio_param_cov} the cosmological parameter correlation matrix in the \Eseparate\ case.

We also calculate the figure of merit (FoM) for the two cases as follows:
\begin{equation}
    \text{FoM}=\frac{1}{\sqrt{\text{det}\left(\mathcal{F}^{-1}\right)}},
    \label{eq_FOM}
\end{equation}
where $\text{det}\left(\mathcal{F}^{-1}\right)$ is the determinant of the parameter covariance matrix. This FoM is the inverse of the volume space of the parameters, hence the higher this quantity, the tighter the parameter constraints. The six cosmological parameters' FoM in the joint case is $\text{FoM}_{\text{joint}}=1.17\times 10^{15}$, while for the separate case, $\text{FoM}_{\text{sep}}=1.06\times 10^{15}$. The two FoMs differ by 10\%, the separate case having a lower FoM in comparison with the joint case, as expected due to the difference in \fsky. This difference is negligible when comparing with \planck\ 2018 \cite{planck18-6}, with the FoM ratio FoM/FoM$_{\planck}$ decreasing from 1.66 to 1.50 when going from the joint to the separate analysis.

We also test a case where we do not apply the transfer function shown in Fig.~\ref{fig_tf} but instead apply a high-pass filter with a cutoff at $\ell=50$ and set $\ellmin=50$.  $\ell = 50$ corresponds to an angular separation of approximately 4 degrees, which is close to the extent in declination of some fields such as the \textit{Wide} a, b, or c fields. We find the same conclusion regarding the diagonal elements of the bandpower covariance matrix, the parameter covariance matrices, and the FoM.

We conclude that the difference in constraining power between the \Eseparate\ and \Ejoint\ cases is negligible. Given the smallness of the effect both on bandpower error bars and on the \LCDM\ parameters, we do not explore the difference for extended models. 
Note that there is additional information in the \Ejoint\ case if one uses \ellmin\ smaller than 50. However, we do not consider this case, given the difficulty of measuring these modes with a ground-based experiment due to the low-$\ell$ atmospheric noise. We thus establish that analyzing the fields separately as in the \EseparateF\ case is a viable analysis strategy without loss of information.

\subsection{\ext\ forecasts}
\label{sec:results_MCMC}
Since the previous section demonstrated that the \EseparateF\ approach is a viable strategy, in this section we perform forecasts using the 13 likelihoods for TT, TE, and EE built for each of the fields, with the realistic noise levels of each field. For brevity, we will label \EseparateF\ as \ext\ in the rest of this work. We include here the simulated lensing likelihood described in Sec.~\ref{sec_lensing} and we derive realistic forecasts on cosmology from the full constraining power of the SPT-3G dataset by running MCMC using \cobaya\ \citep{torrado21}. We consider our chains to be converged when the Gelman-Rubin \cite{Gelman_Rubin} statistic $R-1$ reaches 0.02.

We first compare the results on \LCDM\ with those of \cite{prabhu24}. We then explore parameter constraints for three models that were selected in \cite{khalife24} which are the case of a varying electron mass \cite{uzan2010}, a varying electron mass in a spatially curved universe \cite{sekiguchi2021}, and the case of early dark energy (EDE) \cite{karwal2016}, the last two still being viable solutions to solve the Hubble tension. 

\subsubsection{\LCDM}
\begin{figure*}
\includegraphics[width=0.9\textwidth]{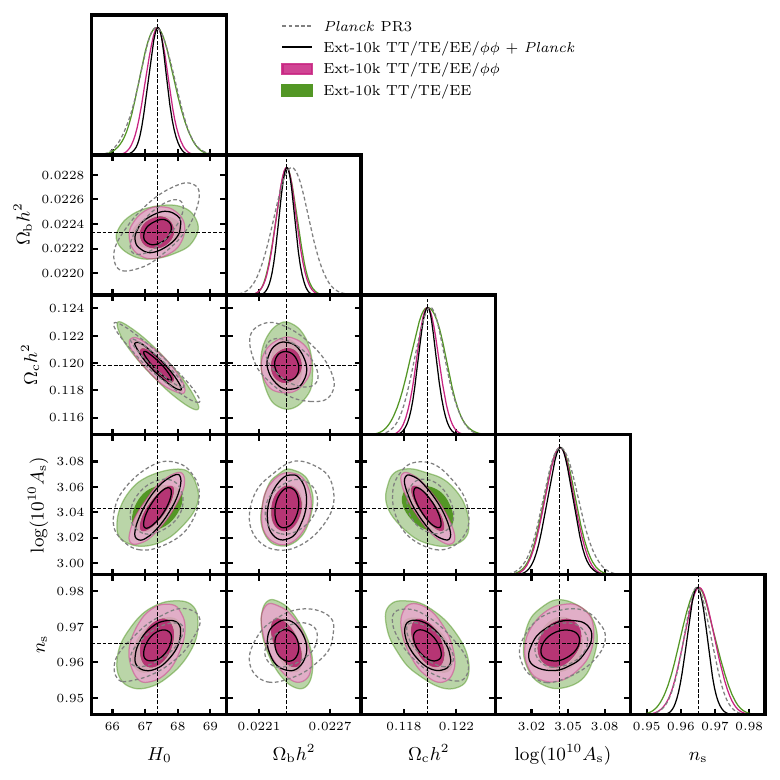}
\caption{SPT-3G \ext\ forecasts of \LCDM\ cosmological parameters. Contours correspond to the 68\% and 95\% confidence levels. We compare our results with \planck\ PR3 TT/TE/EE/\PP\ in gray dashed lines. The green contours show the posterior distributions of the parameters from temperature and polarization combining the 13 SPT-3G TT/TE/EE likelihoods. The pink contours are obtained by adding the five SPT-3G lensing likelihoods. The black line shows the combination of the 18 SPT-3G likelihoods with our mock \planck\ likelihood. The black dashed lines indicate the fiducial values used to produce our simulated bandpowers. SPT-3G \ext\ TT/TE/EE/\PP\ will constrain \LCDM\ cosmological parameters more tightly than \planck\ and almost twice better for $\ombh$.}
\label{fig_LCDM}
\end{figure*}

We first explore constraints with \ext\ on the six cosmological parameters of the \LCDM\ model. We sample over 33 parameters in total, including the six cosmological parameters and the 27 nuisance parameters defined in Sec.~\ref{subsection_model}.  Results are shown in Fig.~\ref{fig_LCDM} and Table~\ref{tab_LCDM}. In Fig.~\ref{fig_LCDM}, we present three SPT-3G posterior distributions corresponding to \ext\ TT/TE/EE, \ext\ TT/TE/EE/\PP, and \ext\ TT/TE/EE/\PP\ + \planck\ which can be compared to \planck\ PR3 constraints that include PR3 lensing. In Table~\ref{tab_LCDM}, we show the associated error bars at the 68\% confidence level and the associated FoM relative to \planck\ PR3. \ext\ TT/TE/EE constraints are obtained by combining our 13 TT/TE/EE likelihoods and using \cosmopower\ to calculate the theory CMB power spectra, while the two other chains are computed using \camb\ for the theory part. SPT-3G \ext\ TT/TE/EE/\PP\ + \planck\ constraints are obtained by combining our 18 SPT-3G \ext\ likelihoods with the \planck\ mock likelihood described in Sec.~\ref{sec_planck}. For the latter we sample over the six cosmological parameters, the 27 SPT nuisance parameters, and the 21 \planck\ nuisance parameters, meaning that we sample over 54 parameters in total. 

\begin{table*}[]
    \centering
    \begin{tabular}{p{2.7cm} c >{\Centering}p{2.5cm} >{\Centering}p{2.5cm} >{\Centering}p{2.5cm} >{\Centering}p{2.5cm}}
    \hline 
    \hline
    Parameter & & \planck\ PR3 & TT/TE/EE & TT/TE/EE/\PP\ & TT/TE/EE/\PP\ + \planck \\
    \hline
    \hline
    \Hubble [\kmsmpc] & $[10^{-1}]$  & 5.4 & 5.1 (1.1) & 3.5 (1.5) & 2.8 (1.9) \\
    \ombh & $[10^{-5}]$ & 15 & 9.0 (1.7) & 8.6 (1.7) & 6.7 (2.2) \\
    \omch & $[10^{-4}]$ & 12 & 13 (0.92) & 8.3 (1.4) & 7.0 (1.7) \\
    \logA & $[10^{-3}]$ & 14 & 12 (1.2) & 12 (1.2) & 10 (1.4) \\
    \ns & $[10^{-3}]$ & 4.2 & 5.1 (0.82) & 4.5 (0.93) & 2.9 (1.4) \\
    \seight & $[10^{-3}]$ & 6.1 & & 3.3 (1.8) & 2.9 (2.1)  \\
    \omegam & $[10^{-3}]$ & 7.4 & & 4.9 (1.5) & 4.1 (1.8)  \\
    \rdrag [\mpc] & $[10^{-1}]$ & 2.9 & & 2.5 (1.2) & 1.9 (1.5)  \\
    \hline
    \hline
    FoM/FoM$_{\planck}$ & & 1 & 2.98 & 9.44 & 31.4 \\
    \hline
    \end{tabular}
    \caption{SPT-3G \ext\ forecasts of \LCDM\ cosmological parameter error bars at the 68\% confidence level. Values in parentheses are the error bar ratios between \planck\ 2018 \cite{planck18-6} and SPT-3G \ext. They show by how much the error bars will be improved with SPT-3G \ext\ in comparison with \planck. We do not show constraints on $\tau$ since they are prior driven. The last row presents the FoM relative to the \planck\ 2018 one. We forecast that SPT-3G \ext\ TT/TE/EE/\PP\ + \planck\ will constrain \Hubble\ and \ombh\ twice as strongly as \planck\ alone.}
    \label{tab_LCDM}
\end{table*}

The SPT-3G \ext\ TT/TE/EE uncertainties on the \Hubble, \ombh, \As, and \omch\ cosmological parameters are similar or lower compared to \planck. In particular, the \ext\ TT/TE/EE \ombh\ error bar is 1.7 times smaller than the one from \planck. We also notice the importance of CMB lensing in tightening the parameter constraints, in particular for \Hubble\ and \omch, for which the constraints are improved by more than 30\% when adding the lensing. SPT-3G \ext\ TT/TE/EE/\PP\ will constrain \LCDM\ cosmological parameters more tightly than \planck\ and almost twice better for \ombh. Finally, adding \planck\ further contributes to tightening \LCDM\ parameter constraints, improving all of them by more than a factor of 1.4 and by more than a factor of 1.9 for \Hubble\ and \ombh\ in comparison with \planck\ alone.

\begin{figure}
\includegraphics[width=0.45\textwidth]{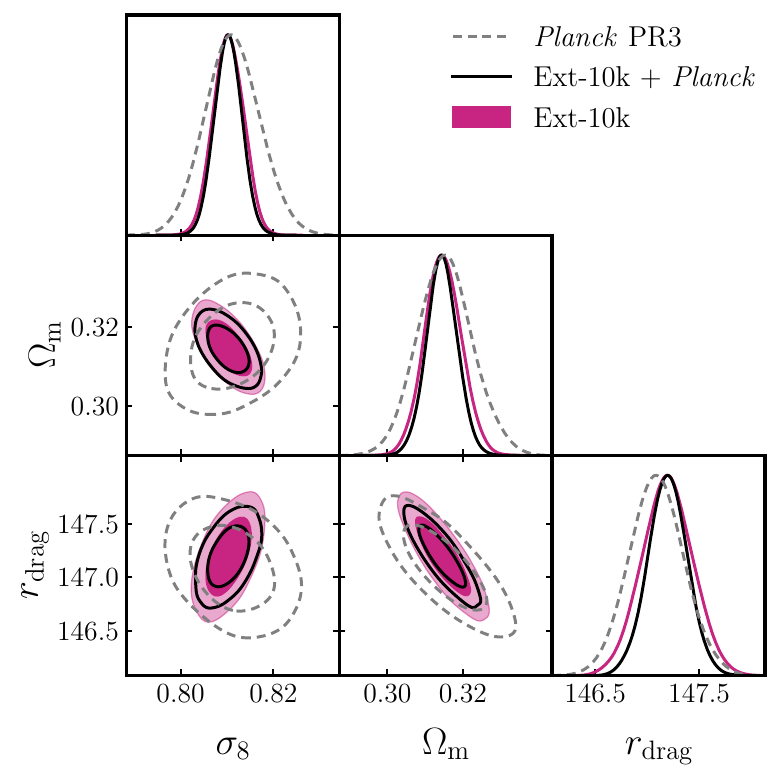}
\caption{SPT-3G \ext\ forecasts of \LCDM\ cosmological parameters \seight, \omegam, and \rdrag. Contours correspond to the 68\% and 95\% confidence levels. We compare our results with \planck\ PR3 TT/TE/EE/\PP\ in gray dashed lines. The pink contours show the parameter posterior distribution from temperature, polarization, and lensing combining the 13 SPT-3G TT/TE/EE and five SPT-3G \PP\ likelihoods. The black line shows the combination of the SPT-3G \ext\ TT/TE/EE/\PP\ with our mock \planck\ likelihood.}
\label{fig_LCDM_sigma8}
\end{figure}

We present additional constraints from SPT-3G \ext\ TT/TE/EE/\PP\ alone or combined with \planck\ on the amplitude of matter density perturbations today \seight, the matter density \omegam, and the sound horizon at the drag epoch \rdrag. Posterior distributions for \ext\ TT/TE/EE/\PP\ and \ext\ TT/TE/EE/\PP\ + \planck\ are presented in Fig.~\ref{fig_LCDM_sigma8}. 
We forecast that SPT-3G \ext\ TT/TE/EE/\PP\ alone will constrain \seight\ almost twice better than \planck, and that its combination with \planck\ will improve \omegam\ constraints by more than a factor of 1.8.

In computing these constraints, we use only one set of parameters to describe the Galactic dust. However, since the \textit{Wide} and \textit{Summer} fields have different locations on the sky, some of them being much closer to the Galactic plane than others, we will be considering 13 different dust amplitudes in the real analysis. Hence, we perform a test in temperature and polarization with Fisher matrices to probe whether having 13 different sets of parameters to capture the potentially different levels of contamination of each field significantly degrades the cosmological parameter constraints. We forecast constraints on parameters with one single Galactic dust amplitude in temperature and with 13 different amplitudes, each of them having the same prior as the one for the single amplitude $A^{\text{TT},\text{Dust}}_{80}$, which is $\mathcal{N}\sim(1.88,0.48)$ from \cite{balkenhol23}. The relative differences between the parameter covariance matrices computed in the two cases are of the order of $\sim 0.002\%$ in the diagonal. Increasing the standard deviation of the prior for the 13 amplitudes to 100 or 1000 does not affect the result. This small difference indicates that our results are robust to dust modeling. 

Next, we check for the consistency of our results and the ones of a previous work presented in \cite{prabhu24}. The latter did not sample over nuisance parameters, hence we also fix them to do this comparison. Fixing the nuisance parameters instead of varying them only changes the error bars of \Hubble, \ombh, and \ns\ by a small amount, with $\sigma(\Hubble)$ changing from $0.35$ to $0.32\,\kmsmpc$, $\sigma(\ombh)$ changing from $8.6\times10^{-5}$ to $8.2\times10^{-5}$, and $\sigma(\ns)$ changing from $4.5\times10^{-3}$ to $4.3\times10^{-3}$. We find that our results are consistent across parameters to within 15\%, which could be explained by some analysis differences. For instance, a difference in the error bars is expected due to the different ways of computing the covariance matrix. In this work we use realistic analytical covariance matrices, while the forecasts shown in \cite{prabhu24} are obtained with the MUSE method \citep{millea22}, a forward simulation-based approach. According to \cite{prabhu24}, the two approaches should be consistent at the 10\% level. Moreover, the $\tau$ prior used in \cite{prabhu24} [$\mathcal{N}\sim(0.0544,0.007)$] is 5\% tighter than ours and the way the lensing is implemented in our analysis is also different since we build separate lensing mock likelihoods, while they build joint TT/TE/EE/\PP\ bandpower covariance matrices. We thus conclude that these small differences can account for the 15\% discrepancies in the two works. Finally, when combining our SPT-3G \ext\ likelihoods with the \planck\ mock likelihood described in Sec.~\ref{sec_planck} to obtain final constraints on \LCDM\ we find that our results are consistent at the 10\% level for \ombh\ and at the 2\% level for the other parameters with those of \cite{prabhu24}. 
Since the two works use different ways of combining with \planck\ and marginalizing over the nuisance parameters, we consider a 10\% difference to be reasonable. In particular, in this work we use a $\tau$ prior instead of the \planck\ low-$\ell$ EE likelihood. We conclude that our forecasts are compatible with those from \cite{prabhu24} and consolidate them with a more realistic patch-based analysis with a foreground marginalization.

We also check that, with our forecasting pipeline, we recover consistent \LCDM\ constraints with the \textit{Main} field real analysis of two years of observation in 2019 and 2020 \cite{Camphuis25}. For this purpose, we use the appropriate noise power spectra, which include two winter seasons from 2019 to 2020, and build a new NKA covariance matrix (see Sec.~\ref{subsection_covariance} for further details). We perform our forecasts in temperature and polarization, and we obtain constraints on \LCDM\ cosmological parameters that are between 15\% and 25\% tighter than \cite{Camphuis25}. A second test with the addition of the marginalization over the temperature and polarization calibration parameters only increases the cosmological parameter constraints by 2\%, 5\%, and 7\% for \Hubble, \omch, and \logA, respectively, which does not fully explain the difference. This difference in the error bars is expected due to additional effects taken into account in the real analysis. For instance, the polarized beams model, described in Appendix~B2 of \cite{Camphuis25}, significantly degrades all parameter constraints, such as the \ns\ error bar which is widened by 20\%. The contribution of gravitational lensing to the covariance matrix also increases the cosmological parameter error bars by 10\%. We do not include the polarized beams model in this work since it was introduced in the real analysis after we started our forecasts. Moreover, a direct measurement of the polarized beams in the coming years should allow for the recovery of these tighter constraints. 

We notice a small shift in our \ext\ TT/TE/EE/\PP\ + \planck\ posterior distribution for \ns\ and \Hubble\ mean values compared to the fiducial values. This small shift is due to the fact that for two of the \planck\ nuisance parameters, $A^{\text{kSZ}}$ and $\xi^{\text{tSZ}\times\text{CIB}}$, the posterior distributions are cut by the lower boundary of the uniform prior, which forces these amplitudes to be positive.
The posterior distribution is thus sampled in an asymmetric way with respect to the peak of the distribution. This introduces a bias in the final distribution of other nuisance and cosmological parameters which are correlated with them. However, the shift is small and should not affect the computation of the error bars on cosmological parameters, which are the quantities of interest in this work.  

Our constraints are obtained with a $\tau$ prior of $\mathcal{N}\sim(0.054,0.0074)$, motivated by \planck\ 2018 results \cite{planck18-6}. Removing this prior widens the constraints on $\tau$, with $\sigma(\tau)$ changing from 0.0069 to 0.018 for \ext\ TT/TE/EE/\PP, and from 0.0061 to 0.011 for \ext\ TT/TE/EE/\PP\ + \planck. This shows that SPT-3G alone will not provide tight constraints on $\tau$ due to the inherent difficulty of measuring large scales with ground-based experiments, these scales being crucial in polarization to constrain this parameter.


\subsubsection{Extended models}
In this section, we use our realistic SPT-3G \ext\ TT/TE/EE and \PP\ mock likelihoods to explore different extended models: two models with a varying electron mass and a model of EDE. SPT-3G has the potential to provide crucial insights into these models due to its precise measurement of small scales in polarization, particularly its measurement of the damping tail. The variation of the electron mass impacts recombination and its duration, which directly affects the damping scale, while EDE impacts the expansion rate before recombination, thus affecting the recombination length. We use the same assumptions and details described in \cite{khalife24,khalife25}. Table~\ref{tab_extended_models_detail} shows constraints on parameters for SPT-3G \ext\ TT/TE/EE/\PP, SPT-3G \ext\ TT/TE/EE/\PP\ + \planck, and for \planck\ alone.

Note that, despite using lensing information to high $\textit{L}$'s, we are not including nonlinear corrections for extended models. We perform a test and enable nonlinear corrections to be computed with \texttt{HALOFIT} \cite{takahashi12} both in the mock bandpowers and the MCMC for the varying electron mass model. We find similar constraints to those obtained without nonlinear corrections. Accurate nonlinear corrections for some of the extended models considered in this analysis are not available yet in codes such as \texttt{HALOFIT} and \texttt{HMCode} \cite{mead15}. However, for models that mainly impact (pre-) recombination physics like the ones considered in this paper, the impact on these corrections should be small.

\textit{\textbf{Varying electron mass.}} We test the varying electron mass model either in a spatially flat universe ($m_{\text{e}}$), where the curvature density parameter \curv\ is fixed to zero, or in a spatially curved space ($m_{\text{e}} + \curv$), where it is allowed to freely vary. In the first case, we sample over seven cosmological parameters: the six common \LCDM\ parameters, and the ratio of the early electron mass and the late electron mass $m_{\text{e,early}}/m_{\text{e,late}}$. In the second case, we sample over eight cosmological parameters by adding the curvature of the universe \curv. We use the model in which the mass of the electron transitions from a value $m_{\text{e,early}}$ to $m_{\text{e,late}}$ at redshift $z=50$, with $m_{\text{e,late}}$ corresponding to the mass of the electron measured in laboratories.
We use \class\ to compute the theoretical CMB power spectra of these two first models. 

\begin{figure}
\includegraphics[width=0.45\textwidth]{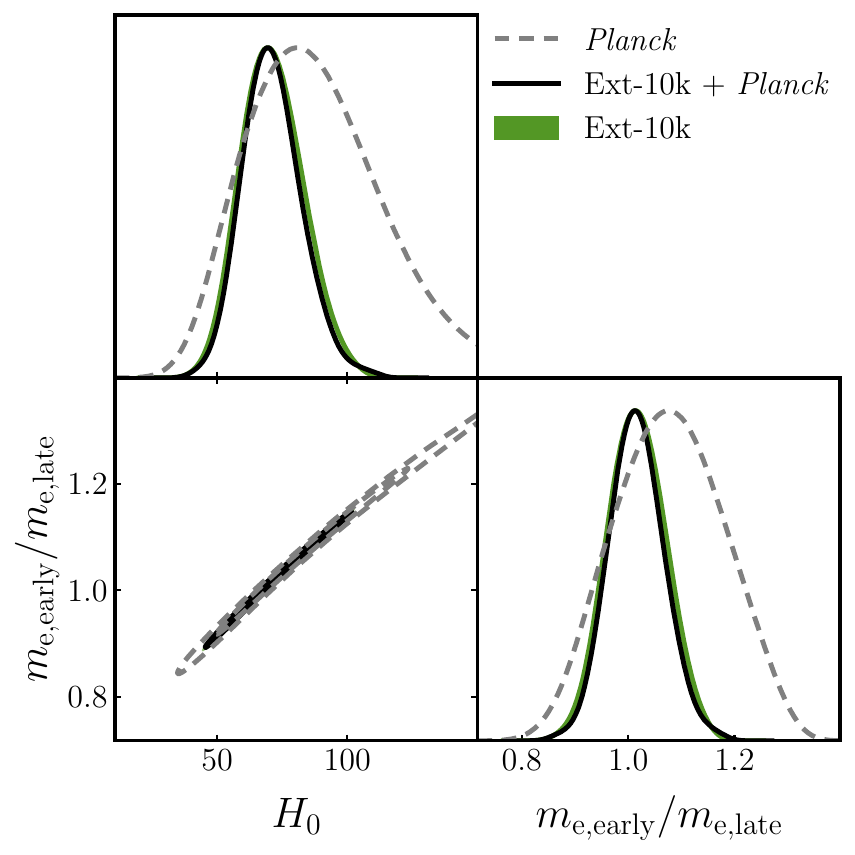}
\caption{SPT-3G \ext\ TT/TE/EE/\PP\ forecasts on the varying electron mass model in a flat universe. Contours correspond to the 68\% and 95\% confidence levels. In this plot, \ext\ refers to \ext\ TT/TE/EE/\PP\ while \planck\ refers to our \planck\ mock likelihood.}
\label{fig_VarMe}
\end{figure}

\begin{figure}
\includegraphics[width=0.45\textwidth]{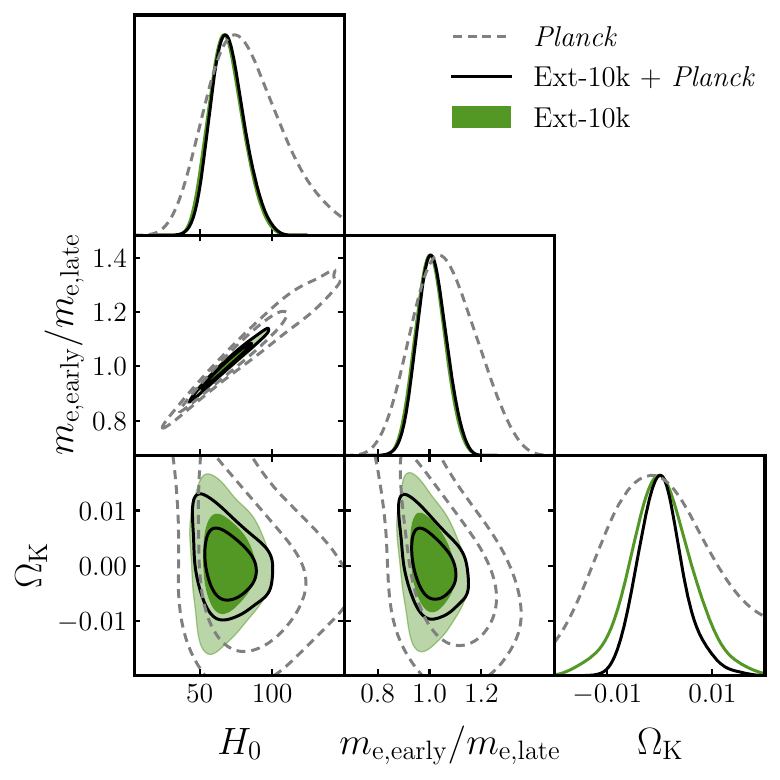}
\caption{SPT-3G \ext\ TT/TE/EE/\PP\ forecasts on the varying electron mass model in a curved universe. Contours correspond to the 68\% and 95\% confidence levels. In this plot, \ext\ refers to \ext\ TT/TE/EE/\PP\ while \planck\ refers to our \planck\ mock likelihood.}
\label{fig_VarMeOmk}
\end{figure}

We show the results for SPT-3G \ext\ TT/TE/EE/\PP\ alone and combined with our \planck\ mock likelihood in Fig.~\ref{fig_VarMe} and Fig.~\ref{fig_VarMeOmk} for $m_{\text{e}}$ and $m_{\text{e}} + \curv$, respectively. Similarly to the previous section on \LCDM\ results, we observe a slight difference between the mean value of the parameters obtained from \ext\ TT/TE/EE/\PP\ alone and in combination with \planck, for the same reasons explained there.


Our forecasted SPT-3G \ext\ constraints on \Hubble\ and $m_{\text{e,early}}/m_{\text{e,late}}$ are a factor of 2 tighter than those from \planck\ in the two cases. We also compute the cosmological parameter FoM of these two models following Eq.~(\ref{eq_FOM}). Namely, the FoM is computed for seven and eight parameters for $m_{\text{e}}$ and $m_{\text{e}} + \curv$, respectively. We compare the FoM obtained for \ext\ TT/TE/EE/\PP\ and \ext\ TT/TE/EE/\PP\ + \planck\ with the one for \planck\ only. In the case of a flat universe with variable electron mass we find that the \ext\ TT/TE/EE/\PP\ FoM is 112 times higher than the one from \planck\ and 192 times better when combining with \planck. In the case of varying electron mass in curved space the \ext\ TT/TE/EE/\PP\ FoM is 162 times higher than that from \planck\ alone and 552 times better when combining with \planck. We notice that combining with \planck\ does not significantly shrink the \Hubble\ error bar in those two models in comparison with SPT-3G \ext\ alone. However we can observe the benefit of combining with \planck\ on other parameter constraints, as shown in Table~\ref{tab_extended_models_detail}, where we detail the error bars of all cosmological parameters.

\begin{figure}
\includegraphics[width=0.45\textwidth]{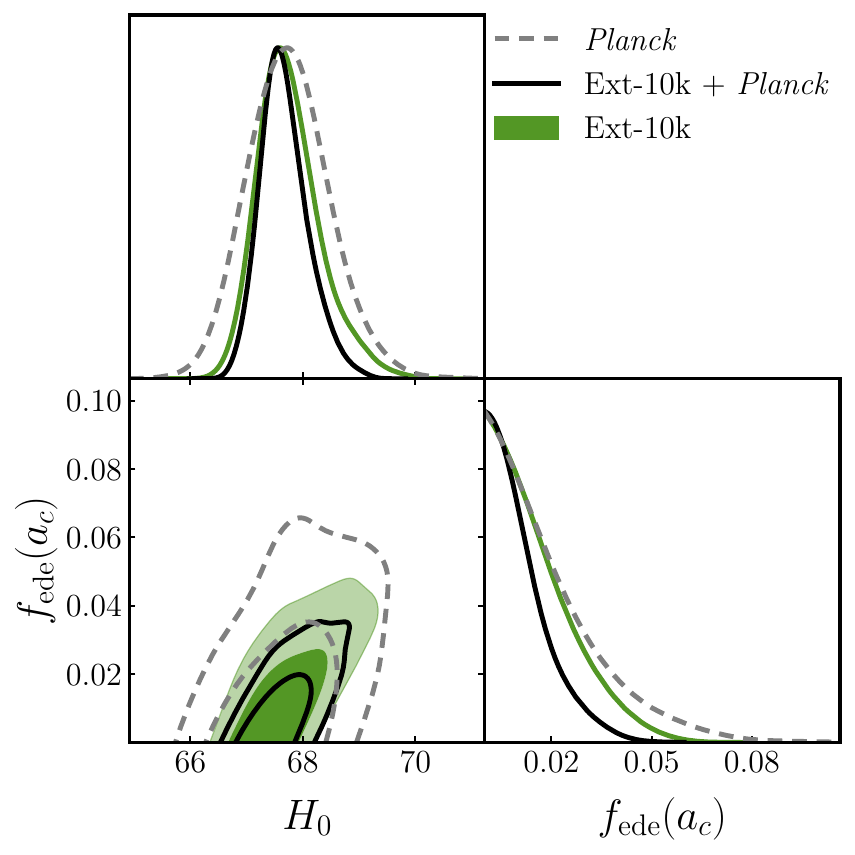}
\caption{SPT-3G \ext\ TT/TE/EE/\PP\ forecasts on the EDE model. Contours correspond to the 68\% and 95\% confidence levels. In this plot, \ext\ refers to \ext\ TT/TE/EE/\PP\ while \planck\ refers to our \planck\ mock likelihood.}
\label{fig_EDE}
\end{figure}

\textit{\textbf{Axion early dark energy.}} For EDE we test the axion early dark energy model \cite{khalife25, poulin19, smith22, McDonough23,cicoli23} using \texttt{AxiCLASS}\footnote{\url{https://github.com/PoulinV/AxiCLASS}} \citep{Smith:2019ihp,Poulin:2018dzj} to compute the theory CMB power spectra in the likelihoods. In this model we sample over nine cosmological parameters: the six \LCDM\ parameters, the initial value of the scalar field $\theta_\mathrm{i}$, the scale factor $a_c$ at a critical time at which the scalar field starts to quickly decay, and the fractional energy of EDE at this time $f_{\text{ede}}(a_c)$. For this model the $R-1$ threshold is increased to $\sim0.05$\footnote{We increase the $R-1$ threshold for this particular model because $\theta_\mathrm{i}$ and $\log_{10}{(a_c)}$ are weakly constrained due to the fact that $f_{\text{ede}}(a_c)$ is compatible with zero. Hence, the sampler will spend a lot of time trying to converge on these two parameters. We base our choice on previous analyses which also use this same value \cite{khalife25,smith22}.} and we also check that the chains have converged by splitting them into two random samples and ensuring they have, visually, the same posterior distribution. We show the results for SPT-3G \ext\ TT/TE/EE/\PP\ alone and combined with our \planck\ mock likelihood in Fig.~\ref{fig_EDE}. 

We forecast that the \Hubble\ constraint from SPT-3G \ext\ is 10\% stronger than that from \planck\ alone and $f_{\text{ede}}(a_c)$ is 1.3 times better constrained with \ext\ than with \planck. We also forecast that the constraints from SPT-3G \ext\ in combination with \planck\ on \Hubble\ and $f_{\text{ede}}(a_c)$ are a factor of 1.6 and 1.8 times tighter than \planck, respectively. The cosmological parameter FoM of this model is computed for nine parameters. We find that \ext\ TT/TE/EE/\PP\ FoM is 18 times higher than that of \planck, and 88 times better when combining with \planck.

\begin{table*}[]
    \centering
    \begin{tabular}{p{2.7cm} | >{\Centering}p{1.5cm} >{\Centering}p{1.5cm} >{\Centering}p{1.5cm} | >{\Centering}p{1.5cm} >{\Centering}p{1.5cm} >{\Centering}p{1.5cm} | >{\Centering}p{1.5cm} >{\Centering}p{1.5cm} >{\Centering}p{1.5cm}}
    \hline
    \hline
    \multirow{2}{*}{Parameter} & \multicolumn{3}{c|}{$m_{\text{e}}$} & \multicolumn{3}{c|}{$m_{\text{e}} + \curv$} & \multicolumn{3}{c}{EDE} \\
    & \planck & \ext & \ext + \planck & \planck & \ext & \ext + \planck & \planck & \ext & \ext + \planck \\
    \hline
    \hline
    \Hubble [\kmsmpc] & $^{+20}_{-30}$ & $^{+10}_{-10}$ & $^{+10}_{-10}$ & $^{+20}_{-30}$ & $^{+10}_{-10}$ & $^{+10}_{-10}$ & $^{+0.67}_{-0.77}$ & $^{+0.42}_{-0.64}$ & $^{+0.34}_{-0.48}$ \\
    \ombh & 0.0024 & 0.0013 & 0.0012 & $^{+0.0025}_{-0.0031}$ & 0.0013 & 0.0013 & 0.00019 & 0.000096 & 0.000085 \\
    \omch & 0.013 & 0.0070 & 0.0067 & $^{+0.014}_{-0.016}$ & 0.0069 & $^{+0.0063}_{-0.0071}$ & $^{+0.0015}_{-0.0019}$ & $^{+0.0011}_{-0.0013}$ & $^{+0.00085}_{-0.0011}$ \\
    \logA & 0.017 & 0.014 & 0.013 & 0.017 & 0.017 & $^{+0.015}_{-0.017}$ & 0.015 & 0.012 & 0.011 \\
    \ns & 0.0052 & 0.0047 & 0.0032 & $^{+0.0057}_{-0.0065}$ & 0.0067 & 0.0043 & 0.0054 & 0.0067 & $^{+0.0033}_{-0.0039}$ \\
    \seight & $^{+0.12}_{-0.089}$ & $^{+0.064}_{-0.057}$ & 0.058 & $^{+0.13}_{-0.11}$ & $^{+0.063}_{-0.057}$ & 0.059 & 0.0087 & 0.0041 & 0.0034 \\
    \omegam & $^{+0.037}_{-0.15}$ & $^{+0.052}_{-0.11}$ & $^{+0.053}_{-0.099}$ & $^{+0.037}_{-0.19}$ & $^{+0.056}_{-0.12}$ & $^{+0.057}_{-0.11}$ & 0.0087 & 0.0055 & 0.0046 \\
    \rdrag [\mpc] & $^{+11}_{-16}$ & $^{+7.4}_{-8.7}$ & $^{+7.2}_{-8.0}$ & $^{+14}_{-18}$ & $^{+7.7}_{-8.9}$ & $^{+7.7}_{-8.7}$ & $^{+0.86}_{-0.40}$ & $^{+0.72}_{-0.36}$ & $^{+0.56}_{-0.27}$ \\
    $m_{\text{e,early}}/m_{\text{e,late}}$ & 0.10 & 0.055 & 0.053 & $^{+0.11}_{-0.13}$ & 0.056 & 0.056 & - & - & - \\
    $\Omega_K$ & - & - & - & $^{+0.0090}_{-0.011}$ & 0.0061 & $^{+0.0037}_{-0.0047}$ & - & - & - \\
    $f_{\text{ede}}(a_c)$ & - & - & - & - & - & - & $<0.0525$ & $<0.0398$ & $<0.0292$\\
    \hline
    \hline
    FoM/FoM$_{\planck}$ & 1 & 112 & 192 & 1 & 162 & 552 & 1 & 18 & 88 \\
    \hline
    \end{tabular}
    \caption{SPT-3G \ext\ TT/TE/EE/\PP\ forecasts of cosmological parameters in the varying electron mass and EDE models. Quoted values are error bars at the 68\% confidence level or upper limits at the 95\% confidence level. We do not show constraints on $\tau$ since they are prior driven. In this table, \ext\ refers to \ext\ TT/TE/EE/\PP\ while \planck\ refers to our \planck\ mock likelihood. The last row presents the FoM relative to the one obtained with our \planck\ mock likelihood alone.}
    \label{tab_extended_models_detail}
\end{table*}


\section{Conclusions}
\label{sec:conclusions}
This work presents and validates the analysis choices for the \ext\ survey of SPT-3G, a high-angular-resolution survey of 10 000 $\sqdeg$ of the CMB from the South Pole.
SPT-3G \ext\ covers 25\% of the total sky and is divided into 13 fields ranging from $-20^\circ$ to $-80^\circ$ in declination. These fields have different characteristics such as different atmospheric contamination, Galactic dust contamination, and noise properties, hence analyzing them independently is our strategy to take into account their particular features. In this paper, we ensure that choosing this analysis strategy does not significantly impact the cosmological parameter constraints. We show that the loss of information from correlations between the fields and the slight decrease of \fsky\ due to the apodization of each individual patch contribute to an increase of less than 3\% of the parameter error bars in \LCDM. 

We then extend the work of \cite{prabhu24} by performing forecasts with a MCMC formalism with a more realistic patch-based analysis and foreground marginalization. While the simulations are more realistic, they still miss a number of components that will be included in the real analysis and that might still affect the constraints. These include the full marginalization over systematic effects or foreground contamination in the lensing reconstruction, which are not considered in this work.
We show that SPT-3G \ext\ TT/TE/EE will constrain some \LCDM\ parameters as strongly as \planck, and we confirm the results of \cite{prabhu24} who showed that SPT-3G \ext\ TT/TE/EE/\PP\ constraints will be tighter than those of \planck\ for the \LCDM\ cosmological parameters. We also expand the grid of extended models treated in \citep{prabhu24} to include certain models with the potential to alleviate the Hubble tension, specifically models with varying electron mass and a model of EDE. 
We find that SPT-3G \ext\ constraints on \Hubble\ will be two times better than those of \planck\ in the varying electron mass case, both in a spatially flat or curved universe. We also forecast that SPT-3G \ext\ will constrain \Hubble\ 10\% stronger than \planck\ in the EDE model. These results show the promising potential of SPT-3G to detect or refute extended models designed to resolve the Hubble tension.\footnote{While proofreading the paper, we noticed that the lensing covariance was wrongly assumed to be 10 times smaller than its actual value. Correcting for this error leads to the results shown here, with a typical increase of 15\% in the error bar of the cosmological parameters in \LCDM, and a reduction of the order of a factor of 10 in the FoM of the extended models compared to \planck. However, this does not affect the overall conclusions of the article and confirms the merits of the \ext\ dataset in constraining a wide range of cosmological models.}

In this work, we do not use galaxy clusters or SZ power spectra, although SPT-3G data will provide additional cosmological constraints using these probes \cite{reichardt21,Raghunathan25,Bocquet25}. In this paper, we are interested in the improvement of constraints derived from CMB data only. In practice, other data are used in combination with CMB to break degeneracies and further constrain extended models, such as BAO with SDSS \cite{beutler11,ross15,alam17,alam21} and DESI \cite{DESI_DR1,desi25}, or type Ia supernovae with Pantheon \cite{Pantheon} and Pantheon+ \cite{Pantheon+}. In conjunction with these probes, SPT-3G \ext\ will provide new insights into our understanding of the Universe.

\section{Acknowledgments}
\label{sec:acknowledgements}
We thank Antony Lewis for scrutinizing the EDE results.
The South Pole Telescope program is supported by the National Science Foundation (NSF) through awards OPP-1852617 and OPP-2332483. Partial support is also provided by the Kavli Institute of Cosmological Physics at the University of Chicago. 
Argonne National Laboratory’s work was supported by the U.S. Department of Energy, Office of High Energy Physics, under contract DE-AC02-06CH11357. 
The UC Davis group acknowledges support from Michael and Ester Vaida. 
Work at the Fermi National Accelerator Laboratory (Fermilab), a U.S. Department of Energy, Office of Science, Office of High Energy Physics HEP User Facility, is managed by Fermi Forward Discovery Group, LLC, acting under Contract No. 89243024CSC000002.
The Melbourne authors acknowledge support from the Australian Research Council’s Discovery Project scheme (No. DP210102386). 
The Paris group has received funding from the European Research Council (ERC) under the European Union’s Horizon 2020 research and innovation program (Grant Agreement No. 101001897), and funding from the Centre National d’Etudes Spatiales. 
The SLAC group is supported in part by the Department of Energy at SLAC National Accelerator Laboratory, under contract DE-AC02-76SF00515.
This work has made use of the Infinity Cluster hosted by Institut d'Astrophysique de Paris. We thank Stephane Rouberol for smoothly running this cluster for us.
The CAPS authors are supported by the Center for AstroPhysical Surveys (CAPS) at the National Center for Supercomputing Applications (NCSA), University of Illinois Urbana-Champaign.  
This work relied on the \texttt{NumPy} library for numerical computations~\citep{numpy}, the \texttt{JAX} library for automatic differentiation and GPU/TPU acceleration~\citep{jax2018github}, and the \texttt{Matplotlib} library for plotting~\citep{matplotlib}.
Posterior sampling analysis and plotting were performed using the \texttt{GetDist} package~\citep{Lewis_getdist}.

\section*{Data availability}
The \ext\ likelihood pipeline is made public \citep{ext10k_pipeline_online}.
\appendix
\section{\polspice\ \texttt{apodizesigma} value determination for the \textit{Wide} fields}
\subsection{The \polspice\ procedure}
\label{apodizesigma_info}

The temperature anisotropies map $\Theta(\hat n)$ is defined from the temperature map $\text{T}(\hat n)$ and the CMB mean temperature T as follows: 
\begin{equation}
    \Theta(\hat n)=\frac{\text{T}(\hat n)-\text{T}}{\text{T}}.
\end{equation}
On the full sky, the temperature anisotropies decomposition into spherical harmonics is
\begin{equation}
    \Theta(\hat n)=\sum_{\ell=0}^{+\infty}\sum_{m=-\ell}^{\ell}a_{\ell m}^{\text{T}}Y_{\ell m}(\hat{n}),
\end{equation}
where the $a_{\ell m}^{T}$ coefficients can be written as
\begin{equation}
    a_{\ell m}^{\text{T}}=\int d\hat{n}\Theta(\hat{n})Y_{\ell m}^{*}(\hat{n}).
\end{equation}
An unbiased estimator of the CMB temperature power spectrum is then calculated from the $a_{\ell m}$'s as follows:
\begin{equation}
    \hat{C}_{\ell}^{\text{TT}}=\frac{1}{2\ell+1}\sum_{m=-\ell}^{\ell}a_{\ell m}^{\text{T}}a^{\text{T}*}_{\ell m}.
\end{equation}
The ensemble average of $\hat{C}_{\ell}$ returns the theoretical spectrum $\langle\hat{C}_{\ell}\rangle =C_{\ell}^{th}$.
However, when not observing the whole sky but only a portion of it we use a mask $W(\hat n)$ which multiplies the temperature anisotropies map so that $\Theta(\hat n)$ becomes $\Theta(\hat n)W(\hat n)$. The $a_{\ell m}$'s become $\tilde{a}_{\ell m}$'s and the associated temperature power spectrum is the pseudopower spectrum $\tilde{C}_{\ell}^{\text{TT}}$,
\begin{equation}
    \begin{array}{l c l} \Theta(\hat{n}) & \longrightarrow & \Theta(\hat{n})W(\hat{n}), \\
a_{\ell m}^{\text{T}} & \longrightarrow& \tilde{a}_{\ell m}^{\text{T}}=\int d\hat{n}\Theta(\hat{n})W(\hat{n})Y_{\ell m}^{*}(\hat{n}), \\ 
\hat{C}_{\ell}^{\text{TT}} & \longrightarrow &  \tilde{C}_{\ell}^{\text{TT}}.
    \end{array}
\end{equation}
Equation~(14) of \cite{hivon02} relates the pseudopower spectrum ensemble average to the theoretical CMB power spectrum as follows:
\begin{equation}
    \langle\tilde{C}_{\ell}\rangle=\sum_{\ell'}M_{\ell\ell'}C^{th}_{\ell'},
    \label{eq_hivon}
\end{equation}
where the matrix $M$ describes the mode-mode coupling induced by the cut sky. 
Inverting this matrix would allow us to recover the theoretical power spectrum. In practice, we do not directly invert it because it is usually ill-conditioned, leading to an inaccurate inversion. In some cases the matrix may even be singular 
due to the fact that it is highly nondiagonal. In this case the inversion is impossible. For these reasons, a regularization is needed and \polspice's solution is to go to real space which is the space of the correlation functions. In real space, Eq.~(\ref{eq_hivon}) is equivalent to the following product:
\begin{equation}
    \langle\tilde{\xi}(\theta)\rangle=w(\theta)\xi^{th}(\theta),
\end{equation}
where $w(\theta)$ is the mask correlation function and $\theta$ is the angular separation in the sky. However, a new problem arises when dividing by $w(\theta)$ since $w(\theta)=0$ when $\theta$ is larger than the mask's physical dimensions. This would induce ringing in the multipole space. To solve this issue, \polspice\ introduces an apodization function $f^{\text{apo}}(\theta)$ which decays smoothly to zero such that
\begin{equation}
    \hat{\xi}(\theta)=\left \{
   \begin{array}{lc}
      \frac{f^{\text{apo}}(\theta)}{w(\theta)}\tilde{\xi}(\theta) & \theta<\theta_{\text{max}}, \cr
      0 & \theta>\theta_{\text{max}},
   \end{array}
   \right .
\end{equation}
where $\theta_{\text{max}}$ is the \texttt{thetamax} parameter chosen by the \polspice\ user and cannot be higher than the masks' physical dimensions. Taking the ensemble average of the pseudocorrelation function gives 
\begin{equation}
    \langle\hat{\xi}(\theta)\rangle=\left \{
   \begin{array}{lc}
      f^{\text{apo}}(\theta){\xi}^{th}(\theta) & \theta<\theta_{\text{max}}, \\
      0 & \theta>\theta_{\text{max}}.
   \end{array}
   \right .
\end{equation}
Going back to the multipole space, the last equation is equivalent to
\begin{equation}
    \langle\hat{C}_{\ell}\rangle=\sum_{\ell'}K_{\ell\ell'}C^{th}_{\ell'},
\end{equation}
where $K$ is the \polspice\ kernel. For the full sky case the kernel is the identity matrix. Otherwise, it is necessary to account for this matrix in the window function during the analysis so that the data vector can be properly compared to the theory CMB power spectrum.

\subsection{The \texttt{apodizesigma} parameter and the \textit{Wide} fields case}
\label{apodizesigma}

We want to find the apodization function $f^{\text{apo}}(\theta)$ such that the kernel $K$ is close to identity. In this work we choose $f^{\text{apo}}(\theta)$ to be a cosine function with $\texttt{apodizetype}=1$ in \polspice,
\begin{equation}
    f^{\text{apo}}(\theta)=\frac{1}{2}\left(1+\cos\left(\frac{\pi\theta}{\sigma^{PS}_\mathrm{apo}}\right)\right),
\end{equation}
where $\sigma^{PS}_\mathrm{apo}$ is called the \texttt{apodizesigma} parameter and is chosen by the \polspice\ user. Larger values of this parameter include larger angular separations. We are interested in determining which values of \texttt{apodizesigma} $\sigma^{PS}_\mathrm{apo}$ and \texttt{thetamax} $\theta_{\text{max}}$ allow us to recover the less biased CMB power spectra.  These parameters can vary from one field to another within the \textit{Main}, \textit{Summer}, and \textit{Wide} fields. However, having the same parameters for all 13 fields of the \ext\ survey implies having the same window function, and this would simplify the analysis.

To estimate these two parameters we start from the \planck\ 2018 best-fit parameters \cite{planck18-6} and we compute one power spectrum $C_{\ell}^{\text{input}}$ with \camb. From this input spectrum we produce 1000 simulations of the CMB map in temperature and polarization with $\texttt{healpy}$ \citep{zonca19,gorski05}. Then \polspice\ multiplies the simulated maps by the mask of the field we consider and computes the 1000 associated power spectra, correcting them with \texttt{apodizesigma} and \texttt{thetamax}. A binning with bin size $\Delta\ell=50$ is applied. Finally, we calculate the mean $\langle\hat{C}_{\ell}\rangle$ and the standard deviation $\sigma_{\ell}$ of these 1000 power spectra and we perform a $\chi^2$ test, where we compare the input spectrum with the mean spectrum as follows:
\begin{equation}
    \chi^2=\sum_{\ell}\left(\frac{\langle\hat{C}_{\ell}\rangle-\sum_{\ell'=0}^{6000}K_{\ell\ell'}C_{\ell'}^{\text{input}}}{\sigma_{\ell}/\sqrt{1000-1}}\right)^2.
\end{equation}
The correlations between multipoles are neglected. The sum is done over an $\ell$ range of $100\leq\ell\leq4000$. We compute the reduced $\chi^2$ from the number of degrees of freedom $N_{\text{dof}}=78$ as $\chi^2_{\text{dof}}=\frac{\chi^2}{N_{\text{dof}}}$.

We focus on the \textit{Wide} fields for the rest of this appendix since a similar study has been previously done for the \textit{Main} and \textit{Summer} fields and showed that $\sigma^{PS}_\mathrm{apo}=30^\circ$ was an appropriate choice for those fields. We use the same apodized masks as the ones described in Sec.~\ref{sec:SPT}. We set $\ellmax=6000$ to produce the simulated maps from the input power spectrum. We set $\texttt{tolerance}=1.5\times 10^{-9}$ in \polspice\ which is the difference between two successive QQ and UU integrals in the convergence process to decorrelate E and B. We impose \texttt{thetamax} = \texttt{apodizesigma} and only vary the \texttt{apodizesigma} value. We tested $\sigma^{PS}_\mathrm{apo}=30^\circ$ and $45^\circ$ for all nine fields because we know these values are reasonable for the \textit{Main} and \textit{Summer} fields. We also tested $\sigma^{PS}_\mathrm{apo}=10^\circ$ and $20^\circ$ for the \textit{Wide} a and b fields to see whether smaller values would improve the reduced $\chi^2$.

Table~\ref{tab_chi2} summarizes the $\chi^2_{\text{dof}}$ values for the nine \textit{Wide} fields. We only show results for $\sigma^{PS}_\mathrm{apo}=30^\circ$ and $45^\circ$ since $\sigma^{PS}_\mathrm{apo}=10^\circ$ and $20^\circ$ do not improve the reduced $\chi^2$ in the \textit{Wide} a and b cases. Moreover, $\sigma^{PS}_\mathrm{apo}=45^\circ$ cannot be applied for \textit{Wide} e, g, h, and i because their physical dimensions are smaller than $45^\circ$. We conclude that $\sigma^{PS}_\mathrm{apo}=30^\circ$ is a reasonable choice for all the \textit{Wide} fields if we want to use the same \polspice\ parameter for all the fields of the \ext\ survey.
\begin{table}[h]
    \centering
    \begin{tabular}{>{\Centering}p{1.5cm} | >{\Centering}p{2cm} | >{\Centering}p{2cm} | >{\Centering}p{2cm}}
    \hline
    \hline
    \textit{Wide} & TT & EE & TE \\
    \hline
    a & 1.06(1.16) & 0.95(0.96) & 0.99(1.27) \\
    b & 1.18(1.13) & 0.94(0.99) & 1.03(1.05) \\
    c & 1.35(1.37) & 1.05(1.02) & 1.27(1.26) \\
    d & 0.96(0.97) & 1.14(1.19) & 0.91(0.88) \\
    e & 0.99 & 1.20 & 1.38 \\
    f & 0.80(0.87) & 0.78(0.76) & 0.86(0.82) \\
    g & 0.91 & 0.91 & 1.16 \\
    h & 0.95 & 1.13 & 0.74 \\
    i & 0.88 & 0.89 & 1.06 \\
    \hline
    \end{tabular}
    \caption{$\chi^2_{\text{dof}}$ values for $\sigma^{PS}_\mathrm{apo}=30^\circ$. Values in parentheses are for $\sigma^{PS}_\mathrm{apo}=45^\circ$. \textit{Wide} e, g, h, and i only have values for $30^\circ$ because their physical size is smaller than $45^\circ$ and \texttt{apodizesigma} cannot have a larger value than the physical size.}
    \label{tab_chi2}
\end{table}

We apply the same methodology for the whole \ext\ patch shown in Fig~\ref{fig_2ways} which covers 23\% of the total sky. We decide to test three different values of \texttt{apodizesigma}: $30^\circ$, $80^\circ$, and $120^\circ$. We find that $\sigma^{PS}_\mathrm{apo}=30^\circ$ and $80^\circ$ give reasonable $\chi^2_{\text{dof}}$ for TT, TE, and EE. We choose to use $\sigma^{PS}_\mathrm{apo}=80^\circ$ for the rest of the \Ejoint\ analysis because it allows us to include larger angular scales than $30^\circ$.

\section{Coadded covariance}
\label{sec:coadded_cov}
In the \Eseparate\ case, we can  calculate the total combined constraining power of the 13 separate fields assuming the coaddition of the separate power spectra after correction of field-specific nuisance and foreground parameters (see Appendix~C4 of \cite{planck15-11}). The coadded $\hat{C}_{\ell}$ covariance matrix can be calculated as
\begin{equation}   
    \text{Cov}(\hat{C}_{\ell}^{\text{coadd}},\hat{C}_{\ell}^{\text{coadd}})=\sum_iw_i^2\text{Cov}(\hat{C}_{\ell}^i,\hat{C}_{\ell}^i),
    \label{eq_cov_coadd}
\end{equation}
where $\hat{C}_{\ell}^{\text{coadd}}$ refers to the coaddition of the power spectra of the different fields,
\begin{equation}
    \hat{C}_{\ell}^{\text{coadd}}=\sum_iw_i\hat{C}_{\ell}^i,
    \label{eq_cl_coadd}
\end{equation}
and $w_i$ is the weight associated to the $i$th field calculated from its \fsky\ value (since we assume uniform noise levels across \textit{Wide} fields),
\begin{equation}
    w_i=\frac{f_{\text{sky}}^i}{\sum_jf_{\text{sky}}^j},
\end{equation}
where $w_i$ is the fraction of the total survey area covered by field i.
This weighting scheme is due to the fact that the variance is proportional to  $1/\fsky$ (see Eq.~(16) of \cite{hivon02}).

\section{Transfer functions}
\label{sec:TF}
Following \cite{Camphuis25} and Hivon \textit{et al.}, in prep, high-pass and low-pass filters are implemented to the real data in order to reduce large-scale atmospheric contamination and small-scale aliasing,\footnote{Small scale aliasing refers to small scales being interpreted as larger scales in the reconstructed signal due to the grid resolution $N_{\text{side}}$ which is too low for these multipoles.} respectively. This can be described by a scalar 
transfer function $F(\ell)$ affecting the measured power spectra, 
\begin{equation}
    \langle C_\ell^{\text{F}}\rangle=F(\ell)\ \langle C_\ell^{\text{U}}\rangle,
\end{equation}
in combination with a rescaling $H(\ell)$ of the covariance matrix diagonal,
\begin{equation}
\text{Cov}(C_\ell^{\text{F}}, C_\ell^{\text{F}})=H(\ell) \text{Cov}(C_\ell^{\text{U}}, C_\ell^{\text{U}}),
\end{equation}
where $C_\ell^{\text{F}}$ and $C_\ell^{\text{U}}$ are the filtered and unfiltered power spectra, respectively. $F(\ell)$ is determined as the ratio of the power spectra of mock-observed maps to the power spectra of the underlying sky maps on the same apodized patch of sky. A set of slow but accurate simulations is used to calibrate the estimate of $F(\ell)$ based on faster but less accurate simulations run in large numbers to reduce the impact of cosmic variance. In this work we assume $H(\ell)=F(\ell)$ since this approximation is accurate at the 10\% level for the \textit{Main} and \textit{Summer} fields based on \cite{Camphuis25} and Hivon \textit{et al.}, in prep. The covariances used in this work are therefore modified from their unfiltered values according to
\begin{equation}
\text{Cov}(C_\ell^{\text{F}}, C_\ell^{\text{F}})=F(\ell) \text{Cov}(C_\ell^{\text{U}}, C_\ell^{\text{U}}),
\end{equation}
where the assumed $F(\ell)$ is shown in Fig.~\ref{fig_tf}. Hence, the relative covariance increases due to the data filtering as follows:
\begin{equation}
\frac{\text{Cov}(C_\ell^{\text{F}}, C_\ell^{\text{F}})}{\langle C_\ell^{\text{F}} \rangle \langle C_\ell^{\text{F}} \rangle} = \frac{1}{F(\ell)} \frac{\text{Cov}(C_\ell^{\text{U}}, C_\ell^{\text{U}})}{\langle C_\ell^{\text{U}}\rangle \langle C_\ell^{\text{U}}\rangle}.
\end{equation}

\clearpage
\bibliography{spt,add_biblio}

\end{document}